\pdfoutput=1
\documentclass[10pt]{article}
\usepackage{wrapfig}
\usepackage{color}
\usepackage{graphicx}
\usepackage[colorlinks=true, pdfstartview=FitV, linkcolor=blue, citecolor=blue,urlcolor=blue]{hyperref}
\def\supp{\mathrm{supp}}

\usepackage{amsbsy}
\usepackage{amssymb}
\def\D{\mathbb D}

\usepackage{verbatim}

\renewcommand{\theequation}{\arabic{section}.\arabic{equation}}

\makeatletter
\@addtoreset{equation}{section}
\makeatother

\newtheorem{theorem}{Theorem}[section]

\newtheorem{exercise}{Exercise}[section]

\newtheorem{lemma}{Lemma}[section]
\newtheorem{remark}{Remark}[section]

\newtheorem{proposition}{Proposition}[section]
\newtheorem{corollary}{Corollary}[section]
\newtheorem{definition}{Definition}[section]
\def\le{\left}
\def\ri{\right}
\def\ds{\displaystyle}

\def\br{\begin{remark}}
\def\er{\end{remark}}
\def\bt{\begin{theorem}}
\def\et{\end{theorem}}
\def\bc{\begin{corollary}}
\def\ec{\end{corollary}}
\def\bx{\begin{examp}\small}
\def\ex{\end{examp}}
\def\bxr{\begin{exercise}\small}
\def\exr{\end{exercise}}
\def\bl{\begin{lemma}}
\def\el{\end{lemma}}
\def\bd{\begin{definition}}
\def\ed{\end{definition}}
\def\bp{\begin{proposition}}
\def\ep{\end{proposition}}
\def\be{\begin{equation}}
\def\ee{\end{equation}}
\def\ov {\overline}

\def\&{\hspace{-15pt}&}
\def\bea{\begin{eqnarray}}
\def\eea{\end{eqnarray}}
\def\beas{\begin{eqnarray*}}
\def\eeas{\end{eqnarray*}}
\def\B{\mathbf B}
\def\A{\mathbf A}
\def \pa{\partial}
\def\C{{\mathbb C}}

\def\R{{\mathbb R}}
\def\N{{\mathbb N}}

\def\Z{{\mathbb Z}}

\def\a{\alpha}
\def\d{\,\mathrm d}

\def\1{{\bf 1}}
\def\wt{\widetilde}

\date{}
\begin{document}
\baselineskip 15pt plus 1pt minus 1pt

\vspace{0.2cm}
\begin{center}
\begin{Large}
\fontfamily{cmss}
\fontsize{17pt}{27pt}
\selectfont
\textbf{
%\begin{itemize}
%\item Outpost colonization of an emerging spectral band
%\item
First colonization of a hard-edge in random matrix theory
%\end{itemize}
%\red{Which title should we choose?}
}
\end{Large}\\
\bigskip
\begin{large} {M.
Bertola}$^{\ddagger,\sharp}$\footnote{Work supported in part by the Natural
    Sciences and Engineering Research Council of Canada
(NSERC).}\footnote{bertola@crm.umontreal.ca}, S. Y. Lee$^{\sharp}$.
\end{large}
\\
\bigskip
\begin{small}
$^{\ddagger}$ {\em Department of Mathematics and
Statistics, Concordia University\\ 1455 de Maisonneuve W., Montr\'eal, Qu\'ebec,
Canada H3G 1M8} \\
$^{\sharp}$ {\em Centre de recherches math\'ematiques\\ Universit\'e\ de
Montr\'eal } \\
\end{small}
\bigskip
{\bf Abstract}
\end{center}
We describe the spectral statistics of the first {\em finite} number
of eigenvalues in a newly-forming band on the hard-edge of the spectrum of a random
Hermitean matrix model. It is found that in a suitable scaling regime, they are described by the same spectral statistics of a finite-size Laguerre-type matrix model.
The method is rigorously based on the Riemann-Hilbert analysis of the
corresponding orthogonal polynomials.

\vspace{0.7cm}

\vskip 15pt
{AMS-MSC2000: 05E35, 15A52}

\tableofcontents

\section{Introduction}
This paper is a sequel and a generalization of our previous paper \cite{BertoLee1}. We consider a Hermitean random matrix model where the spectrum of the matrix is constrained on a pre-determined union of intervals as well as being subject to a (varying) external potential.  In this setting the endpoints of the intervals are called {\bf hard-edges} and---typically---the density of eigenvalues near a hard-edge has power-law of the form $(x-x_0)^{-\frac 12}$.  If---however---the external potential confines the asymptotic spectral density away from the hard-edge, the model is---de facto---independent of the location of the hard-edge.

In our previous paper we described the statistics of the first finitely many eigenvalues that start populating a newly-forming spectral band  and showed that the predictions based on loop equations of \cite{EynardBirth} were in fact correct \cite{MoBirth, Claeys}. Additionally we improved substantially the error estimate, allowing to treat the transitions occurring when the population increases by one. We called the point at which the spectral band is about to emerge the {\em spectral outpost}, picturing the eigenvalues as the first colonies of a large population (living within the main bands).  Pushing the same analogy we could call the hard-edge a {\em port} which has a sea on one side where the colonies cannot live (i.e. where the eigenvalues are forbidden to be).  The present paper is in the same spirit as \cite{BertoLee1}, but we want to describe a similar first-colonization that occurs precisely at a hard-edge of the spectrum (or port).  In the process we will improve some of the ingredients that appeared in \cite{BertoLee1}.

In order to simplify the setup we will simply assume that the model we are describing consists of a model of {\em positive definite} Hermitean matrices $M\in \mathcal H_+$ (here $\mathcal H_+$ denotes the cone of positive definite Hermitean matrices), thereby putting the hard-edge at the origin.  Of course one could as well consider any subset of $\mathcal H$ determined by the requirement that $Sp(M) \subset \bigsqcup I_j$, i.e. the spectrum lies within a pre-determined union of disjoint intervals. Since our analysis is local to one hard-edge there is little loss of generality in restricting ourself to the case of positive matrices. The reader will recognize that all the considerations can be extended without much effort to the most general situation.

We thus consider the matrix model with (unnormalized) probability measure
\be
\d \mu := (\det M)^\alpha {\rm e}^{-\frac N T Tr V(M)}\d M \ ,\ \ \ \alpha > -1 \ ,\ \ M>0.
\ee
The potential $V(x)$ is a (scalar) function with the properties that it is
{\em real-analytic} at $x=0$ (i.e. at the hard-edge),
%on $x\geq -\epsilon$ (for some positive $\epsilon$),
it is bounded from below on $\R_+$ and grows faster than $\ln (1+x^2)$ at infinity.

Following our  idea in \cite{BertoLee1} we will use a simplified setup which dispenses us with the necessity of a complicated double-scaling limit that amounts to a step-wise (infinitesimal) modification of the potential near the hard-edge.%\footnote{In Appendix \ref{App1} we explain how this simplified approach yields in fact identical results to a more usual double-scaling.}

%Suppose that  the mean-field electrostatic potential  is positive on $x>0$ but vanishes as  order $ C_0\,x^\nu$ at $x=0$  ($C_0>0$) but it is strictly positive on a finite interval on the right of the hard-edge. This means that the main spectral band is confined away from the hard-edge and a new band is  ``growing''  at the hard-edge towards the interior of the allowed spectrum.
%Following \cite{BertoLee1} we will add a piecewise constant perturbation of order   $ \ln (N)/N$ to the potential near the hard-edge.

In order to explain the setup in more detail, suppose that the (electrostatic) effective potential \cite{SaffTotik} $\varphi(x) = \frac12 V(x) - g(x) + \frac12\ell$\footnote{The ``$g$--function'' is a rather common object in this area of research and will be explained in due course.} vanishes at the hard-edge  $x=0$ as $C_0 \,x^\nu$ with some $C_0>0$ (the positivity of $C_0$ means that the main spectral band is confined away from the hard-edge).  We then modify the potential (Sect. \ref{SectChemical})  by adding a step-like perturbation of the form
\be\label{epsilonN}
V(x)\to \wt V(x) = V(x) - \le (2T\gamma \varkappa \frac{\ln N}{N}  + \epsilon_N (x) \ri )\chi_J(x)\ ,\qquad \gamma:=\frac 1{\nu}\ .
\ee
Here $\chi_J$ is the characteristic function of a small interval $J$ around the hard-edge. The discontinuous character of the perturbation makes no trouble whatsoever, since the characteristic function could be replaced by any smooth function that is identically $1$ on $J$ and identically vanishing in a slightly larger interval and all the analysis would be identical.
The real parameter $\varkappa$ determines the strength of the perturbation and the constants are crafted for later convenience.
The function $\epsilon_N(x)$, which is independent of $\varkappa$ and or order $N^{-\gamma}$, will be specified in due course.

Due to Dyson's theorem,  instead of studying directly the spectral statistics we focus on the associated orthogonal polynomials and the corresponding {\em Christoffel--Darboux kernel}, in terms of which all correlation functions of the eigenvalues can be written \cite{MehtaBook}. These will be studied with the Deift--Zhou steepest descent method \cite{DKMVZ} based on the formulation of the relevant Riemann--Hilbert problem as in 
\cite{FIK0}.

For $\varkappa<0$ the orthogonal polynomials do not exhibit any peculiar behavior in the large $N$ limit; for positive values of  $\varkappa$ new zeroes of the OPs start appearing near the hard edge. The same transitional phenomenon as $\varkappa$ crosses a half-integer is observed naturally, here as in \cite{BertoLee1}: specifically the normalizations we have chosen are such that
in the asymptotic regime there are $K$ roots near the hard-edge, where $K$ is the integer nearest to $\varkappa$. Clearly transitions must occur at $\varkappa\in \N + \frac 12$.

At the level of the random--matrix model it will appear that the first $K$ eigenvalues that are growing from the hard-edge start populating the new band subject to the statistic of a {\bf finite--size} $K\times K$ matrix model of positive matrices subject to the ``microscopic'' (denoted by the subscript ``m'' hereafter) measure
\be
\d \mu_{m} := \det (M_{m})^\alpha {\rm e}^{- Tr\ V_{m}(M_{m})}
\ee
where $M_{m}$ is  a $K\times K$ positive definite Hermitean matrix  (microscopic compared to the $N\to\infty$ original model) and $V_{m}(\zeta)$ will be an arbitrary monic  polynomial of degree $\nu$ in $\zeta$---a microscopic coordinate.

Although the main results will be along the same line of \cite{BertoLee1} there are a few interesting differences.
\begin{itemize}
\item The microscopic potential $V_m(x)$ can be directly controlled by $\epsilon_N(x)$ (\ref{epsilonN}).  In \cite{BertoLee1} $V_m(x)$ was only a monomial.
\item The global parametrix requires a new piece called {\em Szeg\"o function}.  It can be written in an arbitrary higher genus using Theta functions.
\item The method used in \cite{BertoLee1} of the ``partial Schlesinger transform" was not completely general but seemed to require some (upper-triangular) structure of the local parametrix.  Here we show that the same method can be applied in the  most generic situation.
\item The above method provides us with a recursive procedure to construct the asymptotic solution up to an arbitrary order of accuracy limited only by the error terms coming from the simple end-points of the spectrum.
\end{itemize}

In the main body of the paper we make the simplifying assumption that the support of the equilibrium measure consists of one interval ({\bf one--cut assumption}) confined to the interior of the  positive real axis.
Of course, this assumption is not crucial to the main result.
In Appendix \ref{multicut} we show (in a somewhat sketchy form) how to generalize to an arbitrary number of cuts.

%: only one technical detail cannot be fully addressed in this general case, and concerns with the improved asymptotic for some of the exceptional values of $\varkappa\in \N+1/2$ and exceptional spectral curves, namely the equivalent of formula \ref{detB} for the multi--cut case and the corresponding sign. The knowledge of the sign of (the imaginary part of)  (\ref{detB}) is necessary to ensure that in particularly exceptional circumstances certain denominators (\ref{FG}) do not vanish.

%The asymptotic analysis of the latter has been solved in great detail in
%\cite{DKMVZ} where the author used the Riemann-Hilbert formulation of
%\cite{FIK0, FIK1, FIK2, FIK3} and the nonlinear steepest descent analysis of \cite{DeiftZhou}.

\section{General setting}

\subsection{Riemann-Hilbert problem of orthogonal polynomial}
The setting of the present paper is basically identical to \cite{DKMVZ}, \cite{BleherIts}, and especially to \cite{BertoLee1}.   

Consider a Hermitean matrix model with measure given by
\be
\label{matrixmeasure}
\frac 1{Z_N} (\det M)^\alpha\, {\rm e}^{-\frac NT {\rm tr} V(M)} {\rm d} M,
\ee
where $Z_N$ is the normalization constant.
The potential $V(x)$ is assumed to be {\bf real-analytic} on
%a uniform strip containing
the half-line $x\geq 0$.
% for some positive $\epsilon$.
In particular $V(x)$ is analytic at $x=0$.

Let $\{p_n(x)|n=0,1,2,...\}$ be the corresponding (monic) OPs that satisfy the following orthogonality
\be\label{orthogonality}
\int_{\R_+}  p_n(x) p_m(x) x^\alpha {\rm e}^{-\frac NT V(x)}\d x = h_n\delta_{nm}.
\ee

The spectral statistics of the model can be computed in terms of the Christoffel-Darboux kernel \cite{MehtaBook}
\be
K(x,x') = \sum_{j=0}^{n-1} \frac {p_j(x)p_j(x')}{h_j} =  \frac {
p_n(x)p_{n-1}(x')- p_{n-1}(x)p_{n}(x')}{h_{n-1}(x-x')}
\ee

The asymptotic analysis hinges on the following characterization for the OPs in terms of the Riemann-Hilbert Problem (RHP) described hereafter.
Define for $z\in \C \setminus \R_+$ the matrix
\be
Y(z):= Y_n(z):= \le[
\begin{array}{cc}
p_n(z) & {\cal C}[p_n](z)\\
\frac {-2i\pi}{h_{n-1}}  p_{n-1}(z) & \frac {-2i\pi}{h_{n-1}}{\cal C}[p_{n-1}](z)
\end{array}
\ri]\ ,\qquad {\cal C}[p](z):= \frac 1 {2i\pi} \int_{\R_+} \frac {p(x) x^\alpha{\rm e}^{-\frac
NT V(x)}\d x}{x-z}.
\label{OPRHP1}
\ee
Here and below we use the coordinate $z$ for the complex plane and $x$ when restricted to the real line.

The above matrix has the following jump-relations and asymptotic behavior that
uniquely characterize it \cite{FIK0, FIK1, FIK2, FIK3} (we drop the explicit dependence on
$n$ for brevity)
\bea
Y_+(x) = Y_{-}(x) \le[
\begin{array}{cc}
1 & x^\alpha {\rm e}^{-\frac NTV(x)}\cr
0&1
\end{array}
\ri]\ \mbox{on }~x\geq 0\ ,\qquad
Y(z) \sim\big(\1 + \mathcal O(z^{-1})\big) \le[\begin{array}{cc}
z^n &0\cr 0&z^{-n}
\end{array}\ri].
\label{OPRHP2}
\eea
Replacing the orthogonality condition (\ref{orthogonality}) by the above jump (and boundary) conditions (\ref{OPRHP2}) we obtain the Riemann-Hilbert problem for the OPs.
Using this setup we especially want to investigate the asymptotics of the OPs as their degree $n:=N+r$ goes to infinity while $r$ being fixed to an integer.

\subsection{$g$-function}

\begin{figure}
\resizebox{0.25\textwidth}{!}{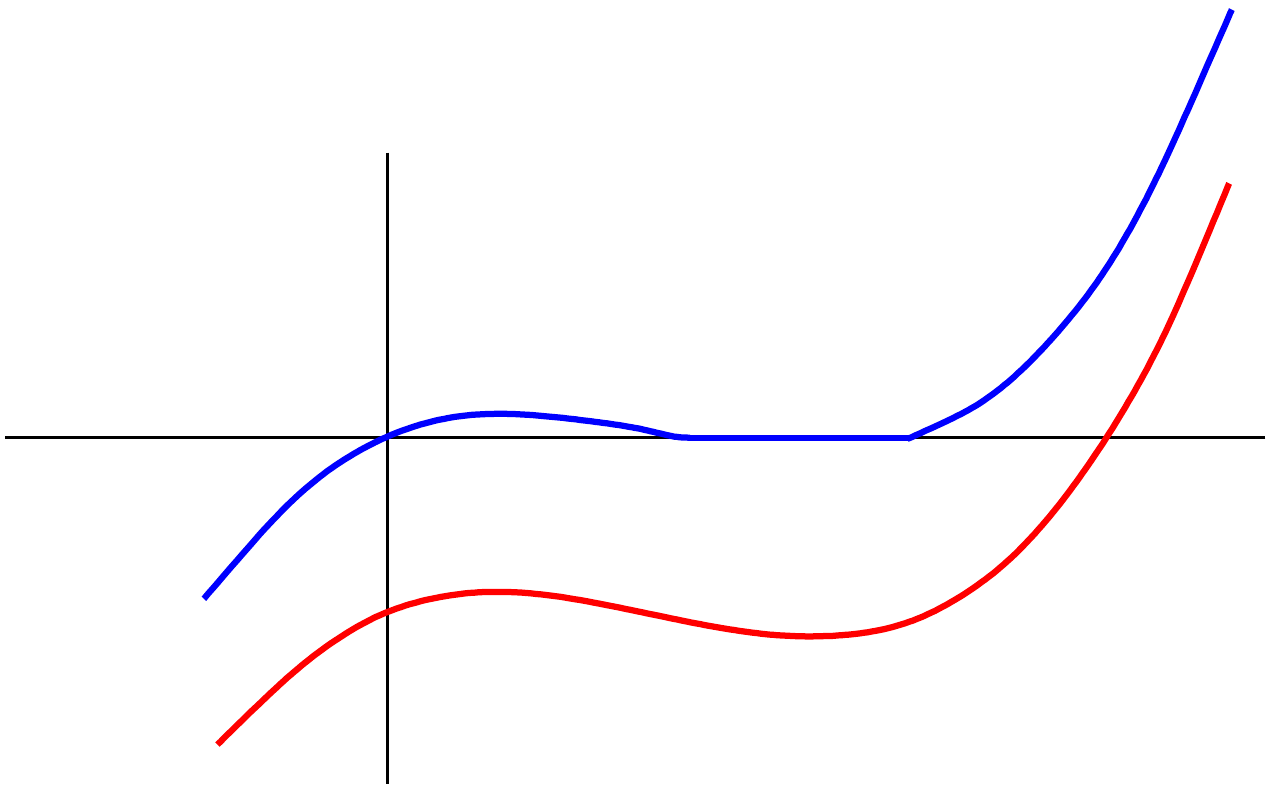}\resizebox{0.25\textwidth}{!}{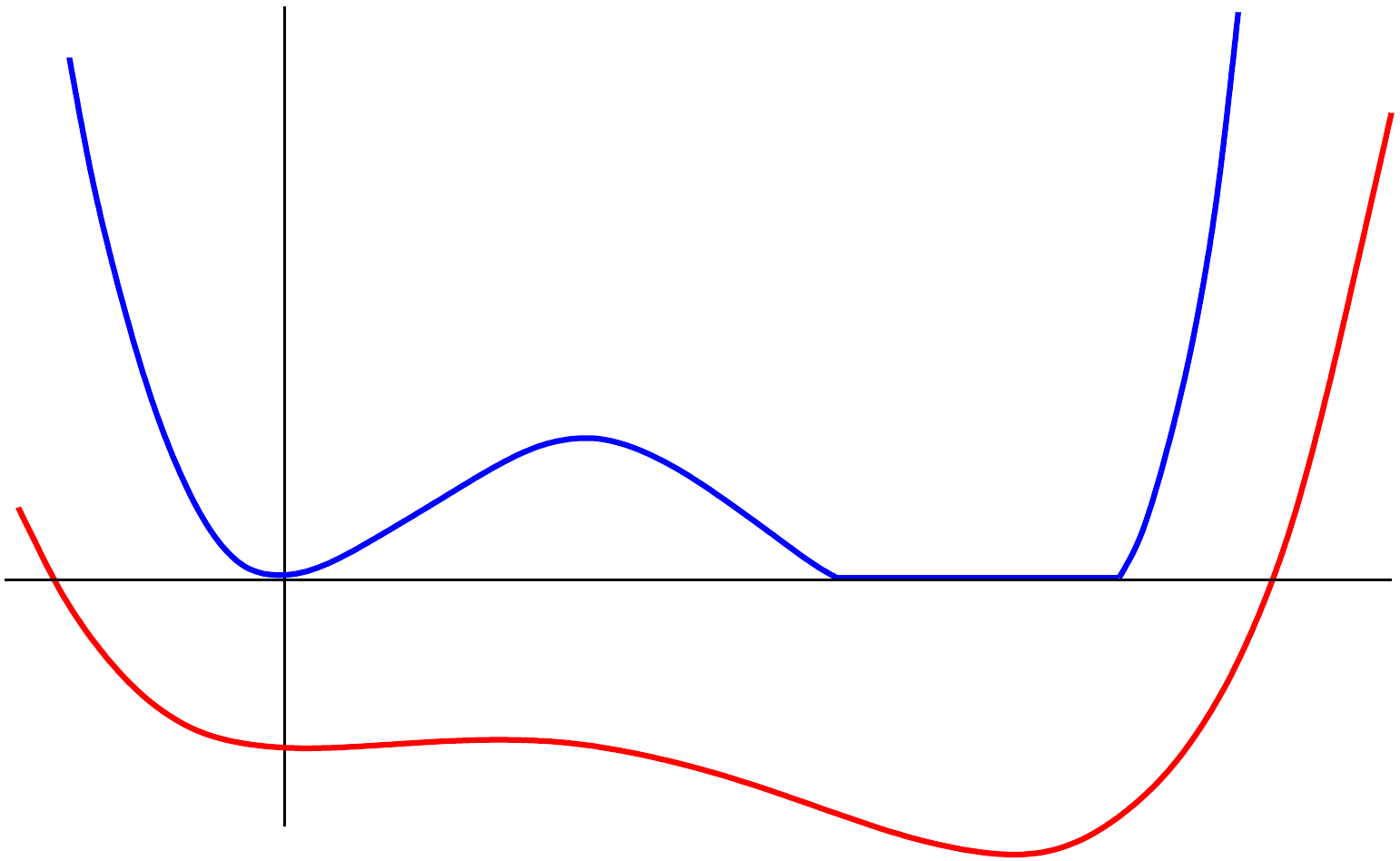}\resizebox{0.25\textwidth}{!}{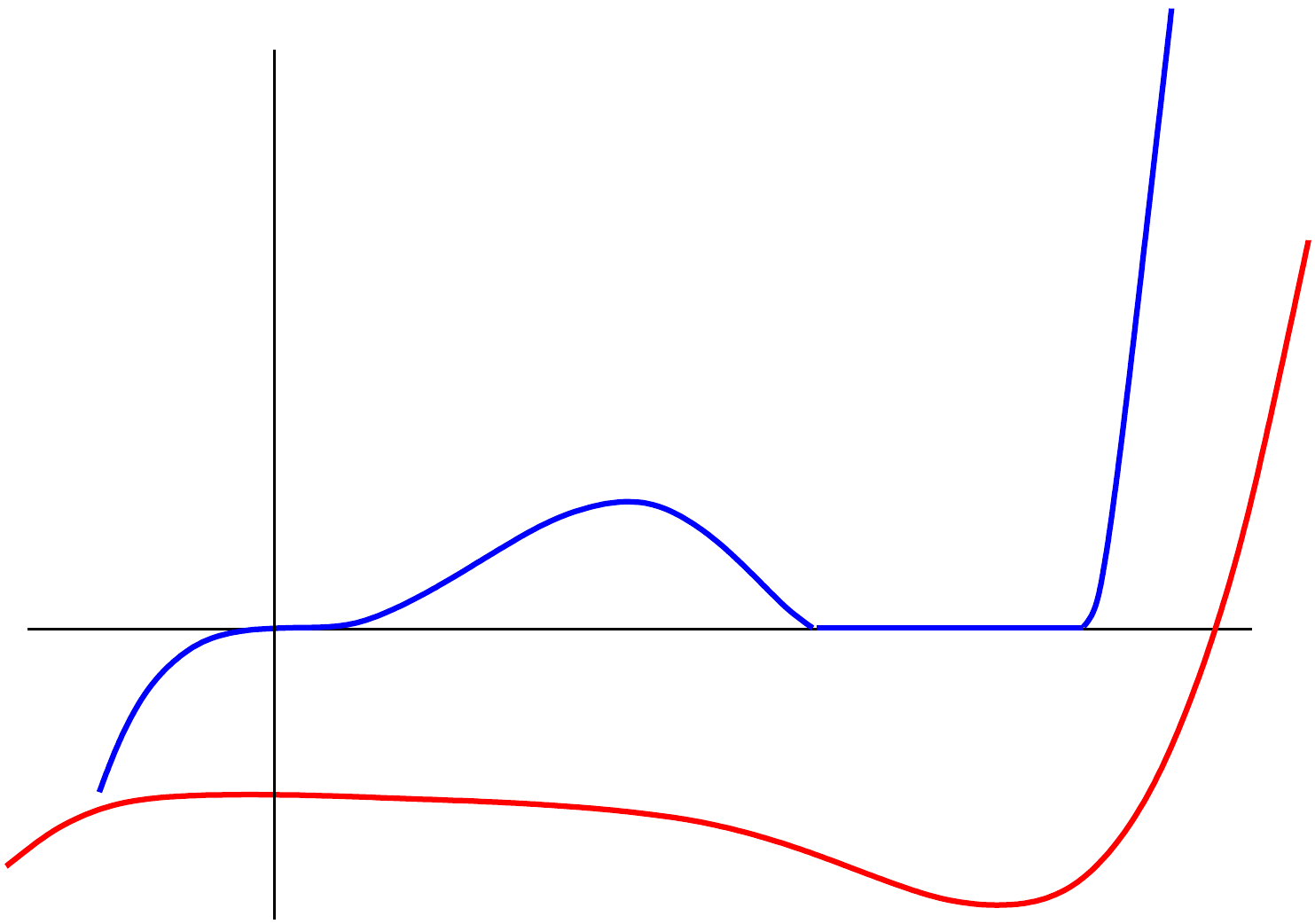}\resizebox{0.25\textwidth}{!}{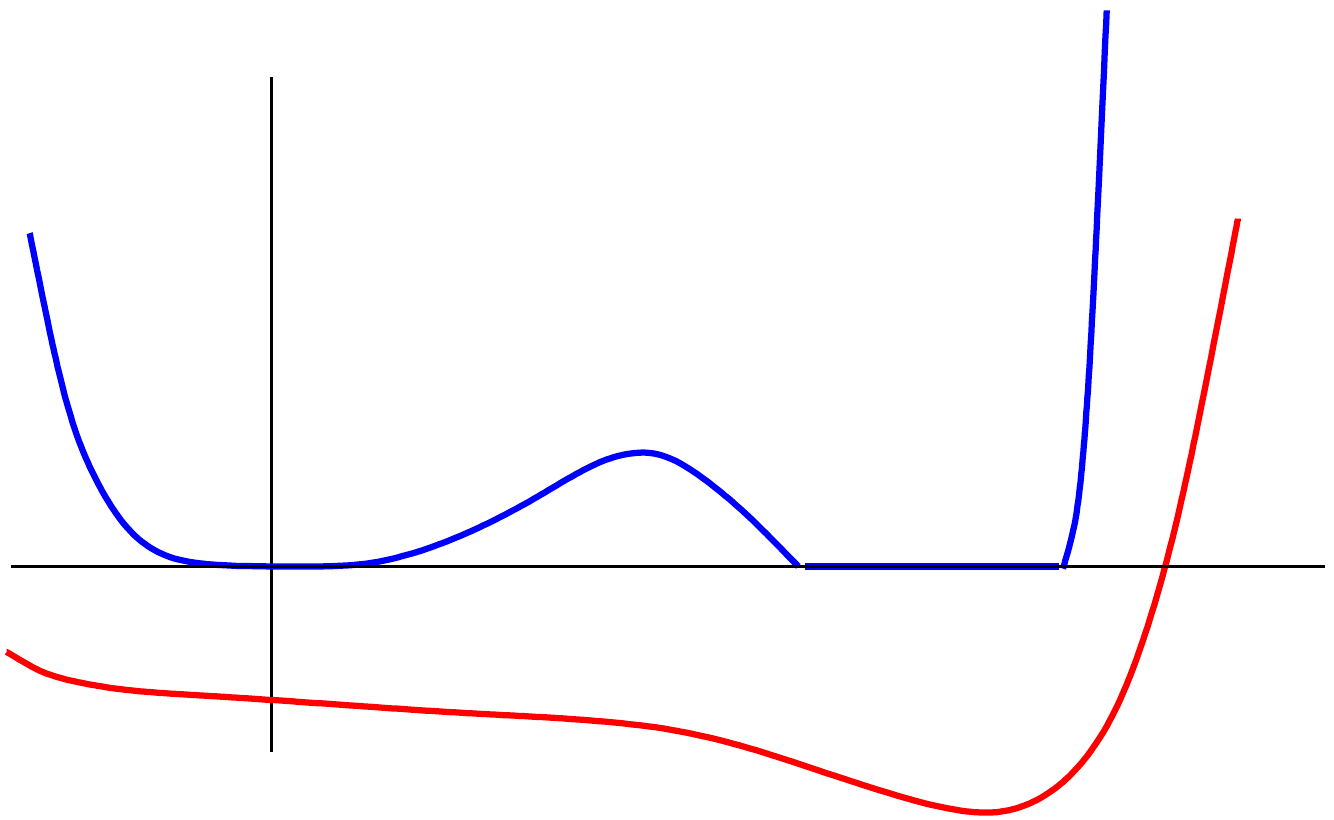}
\caption{Examples of  potentials $ V$  and corresponding effective potentials $\varphi$  for critical hard-edge situations. The effective potential behaves as $x^\nu$ near the hard-edge,  with $\nu =1, 2,3,4$ respectively (from left to right). The plots are numerically accurate, but the vertical axes are scaled differently for $V$ and $\varphi$.}
\label{figeff}
\end{figure}

The asymptotic analysis is based on the construction of the so-called $g$-function;
we recall that, in the simple case where the contour of integration in (\ref{orthogonality}) is the real axis the equilibrium measure is obtained from the solution of a variational problem for a functional over probability measures on the real axis, in the sense of potential theory \cite{SaffTotik}.
Indeed define the {\bf weighted electrostatic energy} \cite{SaffTotik}
\be
\mathcal F[\mu] := \int_{\R_+}  V(x) \d \mu(x)  +  \int_{\R_+} \int_{\R_+}\ln \frac
1{|x-x'|} \d\mu(x)\d\mu(x'),
\ee
where $\d\mu$ is a positive measure supported on the positive real axis with
total mass $T= \int_{\R_+} \d \mu(x)$.

It is known that the functional $\mathcal F$ attains a unique minimum (under
mild assumptions on the growth of $V(x)$ at infinity) at a certain measure $\rho(x)$ that
is called the {\bf equilibrium measure} {\cite{SaffTotik,Deift}}.

 It is also known \cite{McLaughlinDeiftKriecherbauer} that  the
support of the measure $\rho(x)$ consists of a finite union of  disjoint bounded
intervals and that $\rho(x)$ is smooth on the interior of the support.

Taking avail of the equilibrium measure, the $g$-function \cite{Deift} is defined as
\be
g(z):= \int_\R \rho(x) \ln (z-x) \d x=T\ln z +{\cal O}(z^{-1})\label{gfunct},
\ee
where the logarithm must be defined with an appropriate cut extending---say---from the leftmost endpoint of the support of $\rho(x)$ to $+\infty$.

The main properties that enter the steepest descent analysis are the standard
properties of the logarithmic transform. To this end we
 note that the representation (\ref{gfunct}) implies immediately that $\Re g(z)$ is
harmonic away from the support of $\rho$ and  continuous on the whole complex
plane. {The Euler-Lagrange variational equations (equivalent to the optimality of the equilibrium measure $\rho$ \cite{SaffTotik}) can be rephrased in terms of the following conditions for the $g$-function}.
\begin{itemize}
\item for $x\in \R_+$ we have
\be
\Re \varphi(x)\geq 0 \label{ineq}, \quad\varphi(z):=\frac{V(z)}{2}-g(z)+\frac{\ell}{2}, %= \frac{V(z)}{2}-\int \rho(y) \ln (x-y) \d y +\frac{\ell}{2}
\ee
for a suitable real constant $\ell$.  $\Re\varphi$ is the effective potential of the related electrostatic problem.
Since the integration (the {\em conductor} in the terminology of \cite{SaffTotik}) terminates at the origin, the variational equations tell nothing on the sign of $\Re\varphi$ on the left of the hard-edge.
\item The opposite inequality (and hence the equality) holds on the support of $\rho$.  Especially, $\ell$ is chosen such that $\Re\varphi=0$ on the support of $\rho$. (The support of $\rho$ will be called the {\bf cuts} because they form the cuts of the functions $g'(z)$ and $\varphi'(z)$.) {Here and in the previous point, the $g$-function should be understood as the analytic function  defined by its integral representation (\ref{gfunct}) on the simply connected domain obtained by removing a half-line starting e.g. at the leftmost endpoint of the support of $\rho$ and extending towards $\infty$. Then $\Re \varphi(x)$ is actually nothing but the boundary-value $\frac12(\varphi_+(x) + \varphi_-(x)) $} on the half-line.
\item In suitable finite left/right neighborhoods of the cuts, the function $\Re \varphi(z)$, which is
also harmonic in the domain of analyticity of $V(z)$, is {\bf negative}.
\end{itemize}

We note that the effective potential is defined within the domain of analyticity of $V(z)$ and in particular in a left neighborhood of the hard-edge (the origin) due to our assumption that $V(z)$ is analytic at $z=0$.

In particular $\Re \varphi$ is a real analytic function at $x=0$ as long as the support of the equilibrium measure does not contain the hard-edge. If $0$ belongs to the support then it is known \cite{McLaughlinDeiftKriecherbauer}  that---in general---the equilibrium measure $\rho(x)$ has an integrable  singularity of type $1/\sqrt{x}$.

\begin{wrapfigure}{r}{0.35\textwidth}
\resizebox{0.35\textwidth}{!}{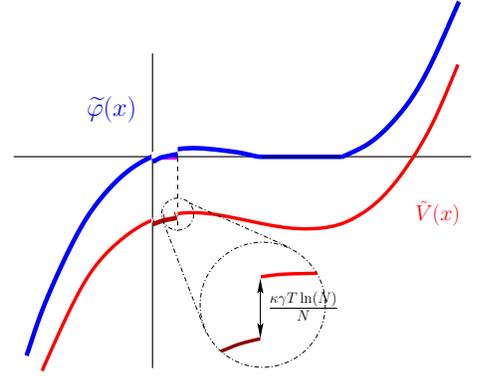}
\caption{The potential $\wt V$ with the chemical potential added, and the corresponding effective potential. In this example $\nu=1$ and the fine-tuning is thus absent.}
\label{figone}
\end{wrapfigure}

We consider the case where $0\not\in\supp (\rho)$ but the inequality (\ref{ineq}) is non strict and fails precisely at the hard-edge. Figure (\ref{figone}) shows the potential and the effective potential.
Taking the cut of $g(z)$ to extend from the leftmost endpoints of the support of $\rho$ to $+\infty$ we see that both $g(z)$ and $\varphi(z)$ are real-analytic functions in a finite neighborhood of the hard-edge $x=0$.
In particular, because of inequality (\ref{ineq}) we must have
\be
\varphi(x) = C_0\,x^\nu (1 + \mathcal O(x))\ ,\ \ C_0>0\ ,\ \ \nu\in \N_+,\label{effvan}
\ee
where $\nu$ can be any {\em positive} integer. The most generic situation is $\nu=1$ when, as we will see, the microscopic ensemble reduces to a Laguerre ensemble.

\subsection{Local coordinate around the hard-edge}

At the hard-edge the effective potential behaves as $\varphi(z)=\frac{1}{2}V(z) - g(z) + \frac{1}{2} \ell \simeq   C_0 z^{\nu}$ with $C_0>0$ (\ref{effvan}).  We define a new conformal parameter $\tilde z$ as follows:
\be
\tilde z := C_0^{-\gamma} \varphi(z)^{\gamma}=z+{\cal O}(z^2)=:z\,{\rm e}^{\eta(z)}.\ee

We define $\mathbb D$ to be a finite open neighborhood around $z=0$ that maps univalently to a {\em disk} centered at $\tilde z=0$ by the above relation and such that $\mathbb D\cap \R_+ = [0,\epsilon)$ does not intersect the support of the equilibrium measure.
We also define the {\bf scaling conformal parameter} by
\be
\label{zetadefine}\zeta:=\left(\frac{2N}{T}\varphi(z)\right)^{\gamma}= (\wt C_0N)^{\gamma}\tilde z\qquad\mbox{where}\quad \wt C_0:=\frac{2C_0}{T}.
\ee
The neighborhood $\mathbb D$ maps to a disk centered at the origin in the $\zeta$-plane whose diameter grows as $N^\gamma$.

 We also define the function $\eta(z)$, which is holomorphic at the origin, by
  \be
 {\rm e}^{\eta(z)}:= \left(\wt C_0 N\right)^{-\gamma} \frac{\zeta(z)}{z} \qquad\mbox{or}\qquad \wt z = z\,{\rm e}^{\eta(z)}.\label{deltacorrection}
 \ee
As can be seen $\eta(z)$ measures the mismatch (up to scaling) of the scaling conformal parameter $\zeta$ with respect to the original coordinate $z$.

%Let us denote by $\zeta_{max} = \sup \zeta (\mathbb D)\cap \R_+$, so that
%\be\zeta_{max} = \mathcal O(N^\gamma).\ee
%We will denote by $\mathcal J= [0,\zeta_{max}]$ the interval in the $\zeta$-plane.

\section{Modifying the setting}

\subsection{Perturbation of the potential}
\label{SectChemical}

Define  $J = [0,\epsilon) = \mathbb D\cap \R_+$;
consider then the following {\bf modified orthogonality relation}:
\bea\label{modOR}
h_{nm} \delta_{nm} =
%\int_{\R_+\setminus J} p_n(x)p_m(x) x^\alpha\,{\rm e}^{-\frac NT V(x)}\d x  +\cr+
\int_{\R_+}p_n(x)p_m(x)\, x^\alpha\, {\rm e}^{-\frac NT \wt V(x)}\d x,\cr
\eea
where we have defined the {\bf exponent of nonregularity} $\gamma:= \frac 1\nu$ and the {\bf perturbed potential}
\bea\label{wtV1}
&&\!\!\!\!\!\!\!\!\!\!\!\!\wt V(x):= V(x) +
\\&&\quad+ \frac T N \bigg( f(\zeta)-(2\varkappa +\alpha)\gamma \ln N -\alpha\gamma\ln\wt C_0  - \alpha \eta(x)\bigg)\chi_J(x). \nonumber
%\\\deg f(\zeta) \leq \nu-1
\eea
The function $f(\zeta)$ is an arbitrarily chosen real polynomial of degree $\nu-1$.

The combersome form of the perturbation is crafted so as to have the simplest local parametrix and so that the parameter  $\varkappa\in{\mathbb R}$ will eventually determine the size of the population of the colony near the hard-edge.

The polynomial $f(\zeta)$ is a {\bf fine-tuning} which allows quite some additional generality: this term did not appear in \cite{BertoLee1} but one could re-read loc.cit. with this fine-tuning in place and the obvious changes while maintaining all the results valid.

\br
For the cases $\nu\geq 2$ the fine-tuning $f(\zeta)$ is in fact quite a strong perturbation since it scales as $N^{1-\gamma}$ which is much stronger than the logarithmic perturbation.  %\red{Later, this will deform the potential of the microscopic ensemble, and effectively resolve the problem of generalization raised at the end of the paper \cite{BertoLee1}.}
\er

%\red{The advantage of the simplified approach using the step function is that it allows us to immediately concentrate on the significant features (the actual RHP) without hindering the analysis into details regarding the appropriate $g$-function.}

% \bea
% Y(z)&\& := Y_{n}(z):= \le[
% \begin{array}{cc}
% p_{n}(z) & \phi_{n}(z)\\
% \frac {2i\pi}{h_{n}}  p_{n-1}(z) & \frac {2i\pi}{h_{n}}\phi_{n-1}(z)
% \end{array}
% \ri]
% \ ,\qquad \phi_n(z):= \frac 1 {2i\pi} \int_\R \frac {p_n(x) {\rm e}^{-\frac
% NT \wt V(x)}\d x}{z-x}\\
% Y_+(x)&\& = Y_{-}(x) \le[
% \begin{array}{cc}
% 1 &  {\rm e}^{-\frac NT V(x)}  N^{2\gamma \varkappa \chi_J(x)}\cr
% 0&1
% \end{array}
% \ri]\ ,\qquad
% Y(z) \sim(\1 + \mathcal O(z^{-1})) \le[\begin{array}{cc}
% z^{n} &0\cr 0&z^{-n}
% \end{array}\ri]
% \eea

\subsection{Normalized and lens-opened RHP}

\begin{wrapfigure}{l}{0.7\textwidth}
\resizebox{0.7\textwidth}{!}{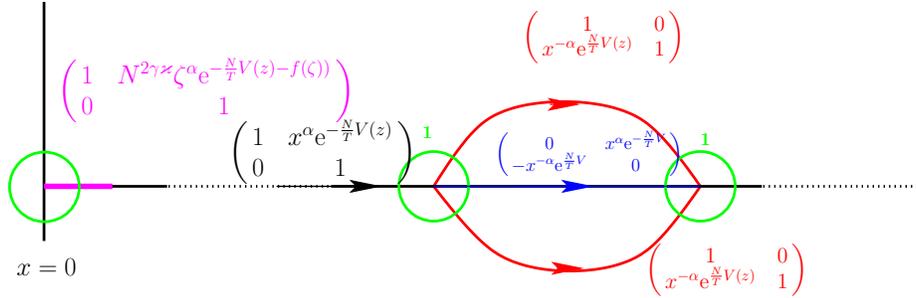}
\caption{\label{figure_Ylens}The jump matrices for $Y$.}
\end{wrapfigure}

In order to streamline the derivation we follow \cite{BertoLee1} and open the lenses {\bf before} normalizing the problem, thus modifying the jumps as shown in figure \ref{figure_Ylens}.  We thus redefine
\bea
&&Y_{\mbox{\scriptsize new}}:=
Y\le[\begin{array}{cc}\scriptstyle1&\scriptstyle0\\\scriptstyle-x^{-\alpha} {\rm e}^{\frac{N}{T} V(x)}&\scriptstyle1\end{array}\ri], \nonumber
\\&&\quad\mbox{ on the upper lip}, \\
&&Y_{\mbox{\scriptsize new}}:=Y\le[\begin{array}{cc}\scriptstyle1&\scriptstyle0\\\scriptstyle x^{-\alpha} {\rm e}^{\frac{N}{T} V(x)}&\scriptstyle1\end{array}\ri], \nonumber
\\&&\quad\mbox{ on the lower lip} .
\eea

For the time being the jumps on the green circles are the identity but later we define separate RHP problems inside the circles.  Then we will call all the RHP inside one of the green disks the {\bf local problem} whereas we call the problem outside the {\bf outer problem}.

After the lens-opening we define
\be\label{Ytilde}
\wt Y(z):=  {\rm e}^{-\frac {N\ell} {2T}\sigma_3}Y(z) {\rm e}^{-\frac N T g(z)\sigma_3}{\rm e}^{\frac {N\ell}{2T}\sigma_3},
\ee
which satisfies a new RHP:
\bea\label{RHPY1}
\wt Y(x)_+ &\&= \wt Y(x)_- \le[
\begin{array}{cc}
\ds {\rm e}^{\frac NT(g_--g_+)} & \ds x^{\alpha}  {\rm e}^{-\frac NT(V- g_+-g_- +\ell)}\left(\wt C_0^{\alpha\gamma}{\rm e}^{ -f(\zeta)+\alpha \eta(x)}N^{(2\varkappa+\alpha)\gamma}\right)^{ \chi_J(x)}\\
0 & \ds {\rm e}^{\frac NT(g_+-g_-)}
\end{array}
\ri]
\\\label{RHPY2}
\wt Y(z)&\& \simeq (\1 + \mathcal O(z^{-1})) z^{ r\sigma_3}\ ,\qquad z\sim \infty
%\\
%\wt Y(z)&\& \simeq (C + \mathcal O(z))  z^{-\varkappa\sigma_3}\ ,\qquad z\sim 0
\eea

\begin{wrapfigure}{l}{0.7\textwidth}
\resizebox{0.7\textwidth}{!}{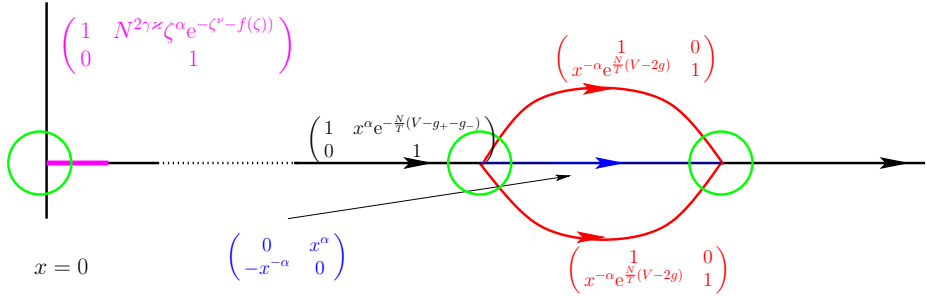}
\caption{\label{figure_Ytildelens} The jump matrices for $\tilde Y$.}
\end{wrapfigure}

For simplicity we assume that there is only one finite band in the spectrum.
This means that $\supp(\rho) = [a,b]$ and that the  the spectral curve $w^2 = (z-a)(z-b)$ is of genus $0$; the generalization to more bands is not conceptually a problem but requires the use of $\Theta$-functions which would make the note quite more technical and long.
Under this assumption, for $x\in{\mathbb R}$,
\bea
g_+(x) &\& = -g_-(x)+V(x)+\ell \ ,\qquad \hbox{ for $x\in\R_+$ on the cut},\\
g_+(x) &\& = g_-(x) - 2i\pi T \ ,\qquad \hbox {for $x\in\R_+$ on the right of the cut},\\
g_+ (x)&\& = g_- (x)\ ,\qquad \hbox {for $x\in\R_+$ on the left of the cut}.
\eea
Everywhere else on the complex plane $g(z)$ is holomorphic.
On account of these properties for the $g$-function the jumps for $\wt Y$ are shown in the figure \ref{figure_Ytildelens}.

In the following the size of the green circles will be fixed to a nonzero value.
In this case the reader can verify that---{\em outside of the green disks}---the jumps on the black and red lines become exponentially close to the identity, and uniformly so  in $L^2\cap L^\infty$.

\subsection{Outer parametrix: Part I}\label{section_outerparametrix}

\begin{wrapfigure}{r}{0.7\textwidth}
\resizebox{0.7\textwidth}{!}{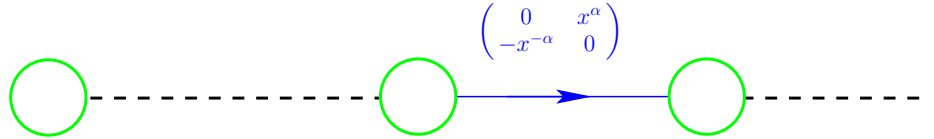}
\caption{Jump matrix for $\Psi_K$.\label{figure_Spinor}}
\end{wrapfigure}

The variational equations (\ref{ineq} and following) imply that all the jumps on the red contours (the lips of the lenses) and on the spectral gaps are exponentially close to the identity jump; after removing them  we are left with the jump matrices as shown in Figure \ref{figure_Spinor}.
This provides the asymptotic RHP whose solution will be called the {\bf outer parametrix}.

The outer parametrix $\Psi_\varkappa$ will be written in the form
\be \Psi_\varkappa(z):=F(z)\,\Psi_K(z),\label{wtPsi}\ee
where $F(z)={\mathbb I}+F_{K,\delta}/z$ is a unimodular (i.e. $\det F(z)\equiv 1$) meromorphic function with a pole only at $z=0$ (at the hard-edge); we will  call this function {\it a partial Schlesinger transform} as in \cite{BertoLee1}.
Remember that the parameter $\varkappa$ appears in the potential  (\ref{wtV1}) and we define the integer $K$ as the nearest integer to $\varkappa$.
Then we define a real number
\be\label{delta} \delta:=\varkappa-K\qquad \mbox{such that}\quad |\delta|\leq \frac12.\ee
The constant matrix $F_{K,\delta}$ will be obtained in section \ref{section_part3} after we introduce the local parametrix.

$\Psi_K$ satisfies the following RHP.
\bea\label{outerRHP1}
&&\Psi_K(z) \simeq  (\1 + \mathcal O(z^{-1})) z^{r\sigma_3},\quad z\sim\infty,
\\\label{outerRHP2}
&&\Psi_K(x)_+  = \Psi_K(x)_-\le[\begin{array}{cc}
0 & x^{\alpha}\\
-x^{-\alpha} & 0
\end{array}
\ri],\quad \mbox{on the cut}.
\\&&
\Psi_K(z) = \mathcal O\big( (z-a)^{-\frac 14}\big),\quad
\Psi_K(z) = \mathcal O\big( (z-b)^{-\frac 14}\big),\quad \label{outerRHP3}
\eea
where $a$ and $b$ are the two turning points.

In fact, these conditions are still met with the left multiplication by $F(z)$, and we need an additional condition to define $\Psi_K$: which is the growth condition at the hard-edge.
\be\label{outerRHP4}
\Psi_K(z) z^{-K\sigma_3} = {\cal O}(1).\ee
The above four conditions (\ref{outerRHP1})(\ref{outerRHP2})(\ref{outerRHP3})(\ref{outerRHP4}) give the Riemann-Hilbert problem for $\Psi_K$.

\subsection{Outer parametrix: Part II (Szeg\"o function)}

%\red{Here we give a specific solution of $\Psi_K$.}

Let $t$ be the uniformizing map of the genus-$0$ Riemann surface.  We let $t_0$ on the $t$-plane to map to the hard-edge on $z$-plane.  For the simplicity of the normalization we choose the location of the cut and the hard-edge in the following way:
\be\label{zt}
z(t) := \frac{b-a}{4}\left(t + \frac 1 t\ri)+\frac{b+a}{2} = \frac{a-b}{4t_0}(t-t_0)\le(\frac 1t-t_0\ri),
\ee
where $a<b$ are the endpoints of the band; in the $t$-plane they correspond to $t=\pm 1$. There are $2$ points in the $t$-plane projecting to $z=0$ namely the two solutions of $z(t)=0$. We denote the one outside the unit circle by $t_0$, the other being $\frac 1 {t_0}$.

\newpage
\begin{wrapfigure}{r}{0.5\textwidth}
\resizebox{0.48\textwidth}{!}{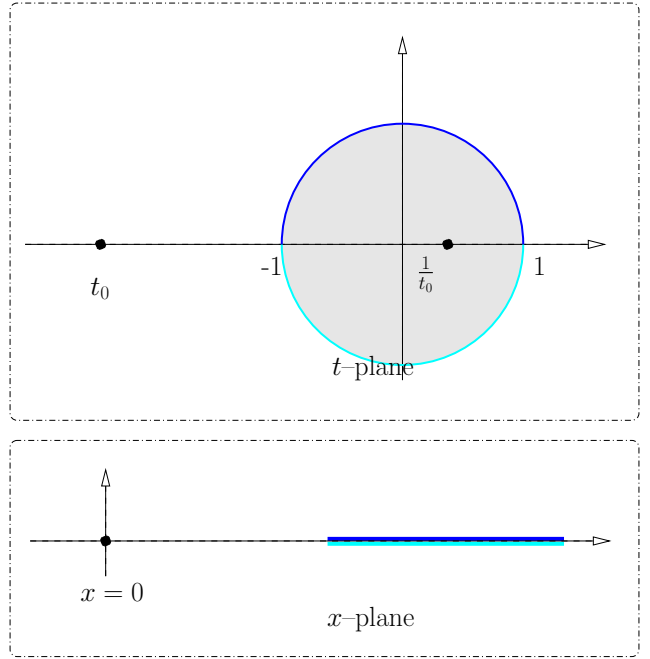}
\caption{The uniformization of the plane sliced along the support of the equilibrium measure. The shaded region is the ``unphysical sheet''.}
\label{fig_tx}
\end{wrapfigure}

In order to solve a RHP with $x$-dependent jumps (\ref{outerRHP2}) we find a function $D(x)$ (usually known as Szeg\"o-function \cite{Vanlessen}) that satisfies
\bea\label{SzegoRHP1}
D_+(x)D_-(x)  = x^\alpha, \ \hbox{ for $x$ on the cut(s)}, \\
\lim_{z\rightarrow\infty}D(z) = D_\infty \quad\mbox{where } D_\infty\in{\mathbb R}_+,\label{SzegoRHP2}
\\ D(x) \mbox{ is analytic and non-zero otherwise}. \label{SzegoRHP3}
\eea
It is immediate to find that
\bea
D(x) = \left(\frac {(t-1/t_0)}{t (t-t_0) } \right)^{\alpha/2}x^{\alpha/2}=\cr
=\left(\frac{b-a}{4}\right)^{\alpha/2}\left(\frac{t-1/t_0}{t}\right)^\alpha,
\eea
where the cuts of the non-integer power go entirely in the ``unphysical sheet'' namely between $t=0$ and $t=1/t_0$.  Also one can see that  $D_\infty=\left(\frac{b-a}{4}\right)^{\alpha/2}$.
In case of several cuts one needs to employ Theta functions and we will give the details in an appendix \ref{App.Szego}.

We now describe a specific solution $\Psi_K$ to the above RHP (\ref{outerRHP1})(\ref{outerRHP2})(\ref{outerRHP3})(\ref{outerRHP4}).

We can write the following solution (a minor modification from \cite{BertoLee1} due to the presence of the Szeg\"o function)
\bea\label{outerK}
\Psi_K(z(t)):= (D_\infty)^{\sigma_3}\, t_0^{K\sigma_3} \frac  { \le(\frac {b-a}4\ri)^{r\sigma_3} }{\sqrt{\frac{4 z'(t)}{b-a} }}\le[
\begin{array}{cc}
t^r & \frac {i} {t^{r+1}}\\
-it^{r-1} & \frac 1 {t^{r}}
\end{array}
\ri]
\le(\frac{t-t_0}{t_0t-  1}\ri)^{K\sigma_3} D^{-\sigma_3}.
\eea
Above, $t$ lies outside the unit circle for $z$ not on the cut (the {\bf physical sheet}), using (\ref{zt}).  We take the branch of the square root such that $\sqrt{\frac{4z'(t)}{b-a}}\sim1$ near $t=\infty$.

\subsection{The local parametrix at the hard-edge}

Within the chosen neighborhood of the hard-edge (inside the green circle)  the RHP for $\wt  Y$ (\ref{Ytilde}) reads
\bea \nonumber
\wt Y_+ &\&=  \wt Y _- \le[
\begin{array}{cc}
1 &z^{\alpha}\left(\frac{2C_0}{T}\right)^{\alpha\gamma} N^{(2 \varkappa+\alpha) \gamma }{\rm e}^{\alpha\eta(z)}{\rm e}^{-\frac N T( V(z) -2 g(z)+\ell) - f(\zeta)} \\
0&1
\end{array}
\ri] = \wt Y _- \le[
\begin{array}{cc}
1 &\zeta ^{\alpha} N^{2 \varkappa \gamma }{\rm e}^{-2\frac N T \varphi(z)    - f(\zeta)} \\
0&1
\end{array}
\ri] = \\
&\& = \wt Y _- \le[
\begin{array}{cc}
1 &\zeta ^{\alpha} N^{2 \varkappa \gamma }{\rm e}^{-V_{m}(\zeta)} \\
0&1
\end{array}
\ri] \qquad\mbox{where}~
 V_{m}(\zeta) := \zeta^{\nu} + f(\zeta).\label{Vmicro}
\eea

We demand that the local parametrix solve the exact same jump within $\mathbb D$.
To define the local parametrix $R_\varkappa(\zeta)$, let us first define $R_K(\zeta)$ that solves the following RHP:
\bea
R_K(\zeta)_+&\&=R_K(\zeta)_- \le[
\begin{array}{cc}
1 &\zeta ^{\alpha}   {\rm e}^{-V_{m}(\zeta)} N^{2 \varkappa \gamma } \\
0&1
\end{array}
\ri]
\qquad\mbox{at} ~\zeta\in \R_+,
\label{Hjump}\\
\label{boundary}
R_K(\zeta) &\& \sim  \zeta^{-K \sigma_3} \mathcal O(1)\qquad \mbox{when}~ \zeta \to 0,\\
R_K(\zeta) &\&\sim \1 +\mathcal O(N^{-\epsilon})
\qquad \mbox{for some } \epsilon>0 ~ \mbox{on}~z \in \partial{\mathbb D}.
\label{newHRHP}
\eea
Here $\epsilon$ is some positive number that will be determined in the subsequent analysis.   Increasing $\epsilon$ leads to a better asymptotics.  The second condition (\ref{boundary}) is equivalent to demanding $\Psi_K R_K$ to be analytic, which implies that $\det R=1$.
Note that the singular behavior at the origin for the local parametrix is on the {\em left}.

Note the appearance of the {\bf microscopic potential} $V_{m}$: in \cite{BertoLee1} this was completely determined by the effective potential and corresponds in the present case to setting the fine-tuning $f(\zeta)$ to zero.  However we realize---a posteriori---that none of the analyisis in \cite{BertoLee1} would be affected had we added the fine-tuning, as we are doing here.

\subsubsection{Case $\varkappa\leq \frac12$}
If $\varkappa\leq\frac12$ then the solution to the RHP (\ref{Hjump})(\ref{boundary})(\ref{newHRHP}) is written as
\be
R_K(\zeta):= \le[
\begin{array}{cc}
1 &  \ds\frac {N^{2\varkappa\gamma } }{2i\pi} \int_{\R_+} \frac {\xi ^\alpha { \rm e}^{-V_{m}(\xi)}\d \xi}{\xi-\zeta}\\[20pt]
0&1
\end{array}
\ri]\qquad\mbox{for}\quad \varkappa\leq\frac12.
\ee

For $\varkappa\leq0$, we may define the local parametrix
\be R_\varkappa(\zeta):=R_K(\zeta)~~~\mbox{and} ~~~F_{K,\delta}=0~~~\mbox{for}~~~\varkappa\leq 0.\label{Rk1}\ee
On $\partial\D$ we have
\be
R_\varkappa(\zeta)\sim \1 + \mathcal O (N^{2\gamma\varkappa-\gamma})\ .
\ee
As we will see, this becomes the error bound of our asymptotics.

When $\varkappa>0$ the error term above will be larger than ${\cal O}(N^{-\gamma})$.
We can lower the error bound using the partial Schlesinger transform as follows.
\begin{equation}\label{Rk2}
R_\varkappa(\zeta):=\left[\begin{array}{cc}
1  &  \ds\frac{1}{\zeta}\frac {N^{2\varkappa\gamma } }{2i\pi} \int_{\R_+} \xi ^\alpha { \rm e}^{-V_{m}(\xi)}\d \xi  \\
 0 & 1     \end{array}\right]R_{K=0}(\zeta)\sim{\bf 1}+{\cal O}(N^{2\varkappa\gamma-2\gamma})\qquad\mbox{for}\quad 0<\varkappa\leq\frac12.
\end{equation}

\subsubsection{Case $\varkappa\geq \frac12$}
We recognize in (\ref{Hjump}) the Riemann-Hilbert problem of the orthogonal polynomials for the weight ${\rm e}^{-V_{m}(\zeta)} \d \zeta$. Specifically, if we denote
by $P_\ell(\zeta)$ the monic orthogonal polynomials that satisfy\footnote{
In the simplest case $\nu=1$ then $f(\zeta)$ is at most an irrelevant constant and the orthogonal polynomials used in the construction of the local parametrix are the Laguerre polynomials with parameter $\alpha$. }
\be
\int_{\R_+}  P_\ell(\xi) P_{\ell'}(\xi)\xi^\alpha  {\rm e}^{-V_{m}(\xi)}\d \xi = \eta_{\ell} \delta_{\ell\ell'}\qquad \eta_\ell>0\ ,
\ee
then the solution of the RHP (\ref{Hjump})(\ref{boundary})(\ref{newHRHP}) is simply given by
\begin{equation}\label{RK}
R_K(\zeta)=\tilde z^{-K\sigma_3}H_K(\zeta),
\end{equation}
where
\bea
H_K(\zeta)&\&:=\wt C_0^{-\gamma K\sigma_3}N^{\gamma(\varkappa-K)\sigma_3}\le[
\begin{array}{cc}
\ds
P_K(\zeta) & \ds \frac 1{2i\pi} \int_{\R_+} \frac{P_K(\xi)\xi^\alpha {\rm e}^{-V_{m}(\xi) }\d \xi}{\xi-\zeta}
\\
\ds \frac {-2i\pi}{\eta_{K-1}}P_{K-1}(\zeta) & \ds\frac {-1}{\eta_{K-1}}  \int_{\R_+} \frac{P_{K-1}(\xi)\xi^\alpha {\rm e}^{-V_{m}(\xi)}\d\xi}{\xi-\zeta}\end{array}
\ri] N ^{-K\gamma \sigma_3}N^{(K-\varkappa )\gamma\sigma_3}\cr
=&\&
N^{\gamma\varkappa\sigma_3}\left(\frac{\wt z}{\zeta}\right)^{K\sigma_3}
\left({\mathbb I}+{\cal O}\left(\frac{1}{\zeta}\right)\right)\zeta^{K\sigma_3}N^{-\gamma\varkappa\sigma_3}.\label{HK}
%\\=&\&
% C_0^{-\varkappa\gamma \sigma_3}\le[
%\begin{array}{cc}\ds\Gamma ^{-K} P_K(\zeta) & \ds \frac { \Gamma^{2\varkappa  - K}} {2i\pi} \int_{\R_+} \frac{P_K(s){\rm e}^{-V_{m}(s)}\d s }{s-\zeta}\\\ds\frac {-2i\pi \Gamma^{K-2\varkappa } }{\eta_{K-1}}P_{K-1}(\zeta) & \ds\frac {-\Gamma^{K} }{\eta_{K-1}}  \int_{\R_+} \frac{P_{K-1}(s){\rm e}^{-V_{m}(s)}\d s}{s-\zeta }\end{array}
%\ri]C_0^{\varkappa\gamma \sigma_3} \ ,\qquad \Gamma:=(C_0 N)^{\gamma}
\eea
It is crucial to point out here that the right multiplier $N^{-\gamma\varkappa \sigma_3}$  is needed to satisfy the correct jump relations (\ref{Hjump}), while the left multiplier $\wt C_0^{-\gamma K\sigma_3}N^{\gamma(\varkappa-K)\sigma_3}$ is needed to restore the boundary condition (\ref{boundary}).

One can satisfy the boundary condition (\ref{newHRHP}) only by choosing $K$ as the closest integer of $\varkappa$; hence $\delta = \varkappa -K\in \le(-\frac 1 2 , \frac 12 \ri)$ as we defined already in (\ref{delta}).

We obtain the following estimate holding {\bf uniformly on the boundary}.
\be
R_K= \1 + \mathcal O(N^{-\gamma+2|\delta|\gamma}),\quad z\in \pa\D\label{errorestimate2}.
\ee

We see that if $\varkappa \in \frac 1 2 + \Z$ then the error term in (\ref{errorestimate2}) does not tend to zero (it is $\mathcal O(1)$).
This is understandable as these values separate regimes where the value of $K$ jumps by one unit and the whole strong asymptotic must changes its form.  A similar problem arose in \cite{EynardBirth}. It was shown in \cite{BertoLee1} and \cite{Claeys} how to overcome the obstacle; there is nothing here that differs substantially from that case and the relevant analysis will be recalled later for the sake of completeness.

\subsubsection{Improved local parametrix}
To define the local parametrix $R_\varkappa$ in this case (and to find the appropriate $F(z)$ appearing in (\ref{wtPsi})),
we first expand $R_K$ for large $\zeta$
\begin{equation}
R_K(\zeta)=\wt z^{-K\sigma_3}\left({\bf 1}+\frac{1}{\zeta}\left[\begin{array}{cc}
\ds a_{K} & \ds -\frac{\eta_K}{2i\pi}\frac{N^{2\gamma\delta}}{\wt C_0^{2\gamma K}} \\
\ds -\frac{2i\pi}{\eta_{K-1}}\frac{\wt C_0^{2\gamma K}}{N^{2\gamma\delta}} & \ds- a_{K}    \end{array}\right]+{\cal O}\left(\frac{1}{\zeta^2}\right)\right)\,\wt z^{K\sigma_3},
\end{equation}
where $a_K$ is defined from $P_K^{\nu}(\zeta)=\zeta^\nu+a_K\zeta^{\nu-1}+\cdots.$

Now we can define the local parametrix $R_\varkappa$ as follows.
\begin{equation}\label{Rk3}
R_\varkappa(\zeta):=\wt z^{-K\sigma_3} \left({\bf 1}-\frac{M_{K,\delta}}{\zeta}\right)H_K(\zeta)\qquad\mbox{for}\quad \varkappa\geq\frac12,
\end{equation}
where the matrix $M_{K,\delta}$ is a {\em nilpotent} matrix defined as follows
\bea
M_{K,\delta}:= \left\{\begin{array}{l}\left[\begin{array}{cc}
 a_{K} & \ds -\frac{\eta_K}{2i\pi}\frac{N^{2\gamma\delta}}{\wt C_0^{2\gamma K}}  \\
 \ds\frac{2i\pi}{\eta_{K}} a^2_K\frac{\wt C_0^{2\gamma K}}{N^{2\gamma\delta}} & - a_{K}    \end{array}\right]\quad\mbox{when $0<\delta\leq\frac12$,}\vspace{0.3cm}\\
 \left[\begin{array}{cc}
 a_{K} & \ds \frac{\eta_{K-1}}{2i\pi}a_K^2\frac{N^{2\gamma\delta}}{\wt C_0^{2K}} \\
 \ds-\frac{2i\pi}{\eta_{K-1}} \frac{\wt C_0^{2\gamma K}}{N^{2\gamma\delta}}& - a_{K}    \end{array}\right]\quad\mbox{when $-\frac12\leq\delta<0$,}\vspace{0.3cm}\\
  \qquad 0\qquad\qquad\qquad\qquad\qquad\qquad\mbox{when $\delta=0$.}
 \end{array}\right.\label{N1}
\eea
These have been chosen such that $R_\varkappa$ gets as close as possible to the identity on the boundary $\partial {\mathbb D}$.
We can verify that
\be R_\varkappa={\mathbb I}+{\cal O}\left(\frac{1}{N^{\gamma \,\min(1+2|\delta|,\, 2-2|\delta|)}}\right)\qquad\mbox{on }\partial {\mathbb D}.\label{wtRKerror}
\ee
Inside the minimum function, $1+2|\delta|$ comes from the $\zeta^{-1}$ term and the $2-2|\delta|$ comes from the $\zeta^{-2}$ term.

\subsection{Outer parametrix: Part III}\label{section_part3}

As the last ingredient to the strong asymptotics, let us calculate $F(z):={\bf 1}+\frac{F_{K,\delta}}{z}$ (\ref{wtPsi}) using the local parametrix obtained in the previous section.
$F(z)$ is uniquely determined by the following analyticity condition at the outpost.
\begin{equation}\label{Fanal}
{\cal O}(1)=F(z) \Psi_K \wt z^{-K\sigma_3} \left({\bf 1}-\frac{M_{K,\delta}}{\zeta}\right)=\left({\bf 1}+\frac{F_{K,\delta}}{z}\right) \Psi_K\, \wt z^{-K\sigma_3} \, \left({\bf 1}-\frac{M_{K,\delta}}{(\wt C_0 N)^{\gamma}} \frac{{\rm e}^{-\eta(z)}}{z}\right).
\end{equation}
This condition guarantees the analyticity of $\Psi_\varkappa R_\varkappa$ at the outpost.

The above condition is equivalent to the two equations below:
\bea\label{z-2}
&z^{-2} \mbox{ term }:&0=F_{K,\delta}\,[\A_K^{(0)},\B_K^{(0)}]\,M_{K,\delta}\ ,\\\label{z-1}
&z^{-1} \mbox{ term }:&0=F_{K,\delta}\,[\A_K^{(0)},\B_K^{(0)}]-
[\A_K^{(0)},\B_K^{(0)}]\, \frac{M_{K,\delta}}{(\wt C_0N)^\gamma}
-F_{K,\delta}\,[\A_K^{(1)},\B_K^{(1)}] \frac{M_{K,\delta}}{(\wt C_0N)^\gamma},
\eea
where the (matrix) coefficients $ [\A_K^{(j)},\B_K^{(j)}]$ are defined by the expansion below.
\be\Psi_K\,\wt z^{-K\sigma_3}\left({\bf 1}-\frac{M_{K,\delta}}{(\wt C_0 N)^{\gamma}} \frac{{\rm e}^{-\eta(z)}-1}{z}\right)=:[\A_K^{(0)},\B_K^{(0)}]+[\A_K^{(1)},\B_K^{(1)}] z+[\A_K^{(2)},\B_K^{(2)}] z^2+\cdots.\ee
Each matrix $[\A_K^{(j)},\B_K^{(j)}]$ for $j=0,1,\cdots,$ depends on $N$ and has a (finite) limit at $N\rightarrow\infty$.

In fact, the equation (\ref{z-2}) is contained in the other one (\ref{z-1}).
Solving the latter equation for $F_{K,\delta}$ and using (\ref{outerK}), we are ready to write down the outer parametrix $\Psi_\varkappa$.
\be\label{wtPsi1}
\Psi_\varkappa(z):= F(z)\,\Psi_K(z)\qquad\mbox{for}\quad \varkappa\geq \frac12,
\ee
with
\begin{equation}\label{F1}
F(z)={\bf 1}+\frac{F_{K,\delta}}{z}\qquad\mbox{where}\quad F_{K,\delta}=[\A_K^{(0)},\B_K^{(0)}]\frac{M_{K,\delta}}{(\wt C_0 N)^\gamma}\left([\A_K^{(0)},\B_K^{(0)}]-[\A_K^{(1)},\B_K^{(1)}]\frac{M_{K,\delta}}{(\wt C_0N)^\gamma}\right)^{-1}\ .
\end{equation}
From the behavior of $M_{K,\delta}$ (\ref{N1}), $F(z)$ is a bounded function of $N$.

\subsection{Asymptotic solution for $\wt Y$ and error analysis}

Collecting all the results we present the following asymptotic solution for $\wt Y$ (\ref{Ytilde}).
\bea\label{wtY}
\wt Y_{\infty}:=\le\{
\begin{array}{cl}
\ds\Psi_\varkappa& \hbox {outside ${\mathbb D}$},\\[5pt]
\ds \Psi_\varkappa \, R_\varkappa & \hbox {inside ${\mathbb D}$}.
%\\[5pt]
%\ds \Psi_K(z) \mathcal A& \hbox { inside the disks around the turning points}
\end{array}
\ri.
\eea
Though we also need separate parametrices around the edges of the cut, we will skip those because they are well known in the literature; for instance, \cite{DKMVZ} (or our previous \cite{BertoLee1}).

%Of course the matrices $\mathcal A(z)$ should be localized at each turning point, but this detail does not differ in any way from known formul\ae\ already present in the literature \cite{DKMVZ}.

\begin{wrapfigure}{r}{0.7\textwidth}
\resizebox{0.7\textwidth}{!}{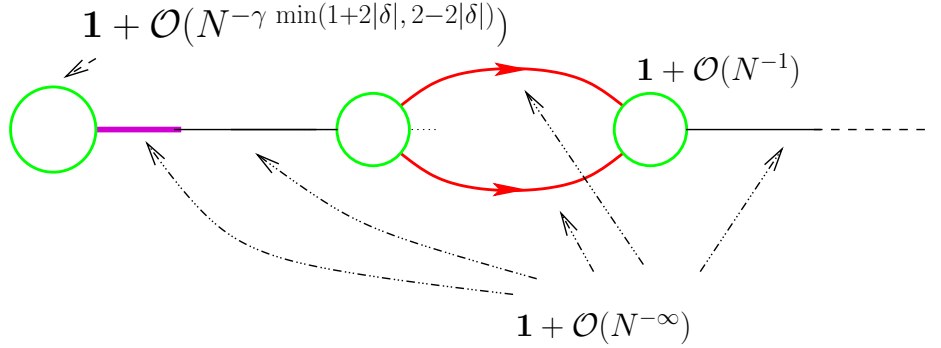}
\caption{\label{fig_Residual} The jumps of ${\mathcal E}$.}
\end{wrapfigure}

To find the error term we define the {\em error matrix} as follows.
\bea
\mathcal E(z):=\wt Y_\infty\wt Y^{-1}\eea
The error matrix solves the residual RHP with the jump matrices as shown in the figure \ref{fig_Residual}.
It follows from the construction that there is no jump inside the green disks, and on the cut.

On the disks around the edges of the cut (with airy parametrix), the jumps converge to the identity with uniform error bounds in $L^2\cap L^\infty$ as (we simply cite the result, for instance from \cite{DKMVZ})
\be
{\mathcal E_+\mathcal E_-^{-1} ={\bf 1}+{\cal O}(N^{-1}).}
\ee

The only non-trivial part is the jump on $\partial{\mathbb D}$ which is given by
\be
{\mathcal E}_+ ({\mathcal E}_-)^{-1}=\Psi_\varkappa R_\varkappa \Psi_\varkappa^{-1}={\bf 1}+{\cal O}\left(\frac{1}{N^{\gamma\,  \min(1+2|\delta|,\, 2-2|\delta|)}}\right)\qquad\mbox{on }\partial {\mathbb D}.
\ee
This is directly obtained from the error bound of $R_\varkappa $ (\ref{wtRKerror}) because $\Psi_\varkappa(z)$ is uniformly bounded on  $\partial {\mathbb D}$ for growing $N$ (\ref{F1}).

A well-known theorem \cite{Deift} guarantees that error matrix itself is bounded by the same error bound, i.e. ${\mathcal E}={\bf 1}+{\cal O}(N^{-\gamma-2\gamma|\delta|})$.  This gives the following error term for the (strong) asymptotics of $\tilde Y$.
\be
\wt Y=\wt Y_{\infty}\left({\bf 1}+{\cal O}\left(\frac{1}{N^{\gamma\,\min(1+2|\delta|,\, 2-2|\delta|)}}\right)\right).
\ee

\section{Roots and kernel at the hard-edge}
\subsection{Case $|\varkappa-K|\neq \frac12$}

Using the strong asymptotics $\wt Y_\infty$ (\ref{wtY}) {\it inside} the disk ${\mathbb D}$, let us look at the behavior of the roots at the hard-edge in detail.

The orthogonal polynomial appears at the (11)-component of the matrix $Y$.  Due to the nature of the transformation: $Y\rightarrow\wt Y$ (\ref{Ytilde})  it is enough to consider $\wt Y$ to study the roots of that polynomial.

Collecting the previous results,
let us write the full expression of $\wt Y_\infty$ to study the asymptotic behavior for large $N$.
\begin{equation}\label{full}
\wt Y_\infty=\Psi_\varkappa\,R_\varkappa=\left({\bf 1}+\frac{F_{K,\delta}}{z}\right)\Psi_K \wt z^{-K\sigma_3}\left({\bf 1}-\frac{M_{K,\delta}}{\zeta}\right)\wt C_0^{-\gamma K\sigma_3}\le[
\begin{array}{cc}
\ds
P_K(\zeta) & *
\\
\ds \frac {-2i\pi}{\eta_{K-1}N^{2\gamma\delta}}P_{K-1}(\zeta)& *\end{array}
\ri] N ^{-K\gamma \sigma_3}.
\end{equation}
The asterisks denote expressions which are irrelevant to the present analysis.

Defining a short-hand notation $\wt M_{K,\delta}:=(\wt C_0 N)^{-\gamma} M_{K,\delta}$, which is of the order ${\cal O}(N^{2|\delta|\gamma-\gamma})$, we may evaluate the leading terms for large $N$ as
\begin{equation}\label{wtYapprox}
\wt Y_\infty=\Big([\A_K^{(0)},\B_K^{(0)}]+\underbrace{{\cal O}(N^{2|\delta|\gamma-\gamma})}_{\zeta\mbox{ independent}}+{\cal O}(N^{-\gamma})\,{\cal O}(\zeta)\Big) \,\wt C_0^{-\gamma K\sigma_3}
N ^{-K\gamma}
\le[
\begin{array}{cc}
\ds
P_K(\zeta) & *
\\
\ds \frac {-2i\pi}{\eta_{K-1}N^{2\gamma\delta}}P_{K-1}(\zeta)& *\end{array}
\ri] .
\end{equation}
The expression above implies that the microscopic zeroes are asymptotically close to those of $P_K(\zeta)$ if $\delta>0$, or $P_{K-1}(\zeta)$ if $\delta<0$ with one stray zero elsewhere.  The details on the subleading behavior of the zeroes are basically equivalent to the description in \cite{BertoLee1}, section 4.4.  Here we summarize the scaling behavior of zeroes when $\varkappa$ is not a half integer.
\begin{itemize}
\item $K<\varkappa<K+\frac12$: All $K$ zeroes are at a distance of order $\mathcal O(N^{-2\gamma\delta})$ (in the $\zeta$-coordinate) from those  of $P_{K}(\zeta)$.
\item $K-\frac12<\varkappa<K$. There are $K$ zeroes in total; $K-1$ zeroes are within a $\mathcal O(N^{-2\gamma|\delta|})$ distance from those of of $P_{K-1}(\zeta)$ and the other zero is running away to the infinity at the rate $\sim N^{2\gamma|\delta|}$ in the $\zeta$-coordinate.   We may locate this zero as
\be\zeta_{\mbox{\scriptsize away}}\approx \frac{2i\pi \wt C_0^{2\gamma K}\B^{(0)}_{1}}{\eta_{K-1}\A^{(0)}_{1}} N^{2\gamma |\delta|} \qquad \mbox{where}\qquad[\A^{(0)}_K,\B^{(0)}_K]=\left[\begin{array}{cc}
 \A^{(0)}_1 & \B^{(0)}_1  \\
\A^{(0)}_2  & \B^{(0)}_2     \end{array}\right].\ee
Using the explicit representation (\ref{outerK}) we obtain
\be \zeta_{\mbox{\scriptsize away}}\approx  - \frac{\pi \wt C_0^{2\gamma K}2^{-2(2K+\alpha)+1}(b-a)^\alpha}{\eta_{K-1} \,t_0^{1+2r}}\left(1-\frac{1}{t_0}\right)^{2\alpha}\left(\frac{t_0^2}{(b-a)(1-t_0^2)^2}\right)^{-2K} N^{2\gamma|\delta|}. \ee
Considering the relative location $a<b$ and $t_0<0$ (see Figure \ref{fig_tx}), the above quantity is positive, which means that the stray zero runs away to $\zeta = +\infty$ (namely towards the main spectral band). Note that in the $x$--coordinate the stray zero actually moves {\em towards} the hard-edge, but at a slower rate than the other ones, namely as $\mathcal O(N^{\gamma(2|\delta|-1)})$.
\item  $\varkappa=K$: There are $K$ zeroes are at a distance of order $\mathcal O(N^{-\gamma})$ from the zeroes of $P_K(\zeta)-\frac{2i\pi \wt C_0^{2\gamma K}\B^{(0)}_{1}}{\eta_{K-1}\A^{(0)}_{1}}  P_{K-1}(\zeta)$.
\end{itemize}

The Christoffel--Darboux kernel (in terms of which all the spectral statistics can be computed) is given by
\bea
K_n(z,z')&\& :=\frac{1}{h_{n-1}}\frac{p_n(z)p_{n-1}(z')-p_{n-1}(z)p_n(z')}{z-z'}
=\frac{1}{2i\pi}\frac{\det\left[\begin{array}{cc}
p_n(z)  & p_n(z')   \\
\frac{-2 i\pi}{h_{n-1}} p_{n-1}(z)  & \frac{-2 i\pi}{h_{n-1}} p_{n-1}(z')     \end{array}\right]
}{z'-z}=
%\cr&\&=\frac{1}{2i\pi}\frac{[\wt Y^{-1}(z')\wt Y(z)]_{21}}{z-z'}{\rm e}^{-\frac NT\le(g(z)+g(z')+\ell\ri)}
\cr&\&=\frac{{\rm e}^{\frac NT\le(g(z)+g(z')-\ell\ri)}}{2i\pi}\frac{1}{z'-z}\det\left[\begin{array}{cc}
 \wt Y_{11}(z) & \wt Y_{11}(z')  \\
 \wt Y_{21}(z) & \wt Y_{21}(z')     \end{array}\right].
\eea
From the above and from (\ref{wtYapprox}) we can evaluate the leading behavior of the kernel in the scaling coordinate $\zeta$.
Using our asymptotics $\wt Y\simeq \Psi_\varkappa R_\varkappa$, we evaluate the kernel in the local coordinate near the hard-edge as
\bea
K_n(\zeta,\zeta')\simeq\frac{\wt C_0^{\gamma}{\rm e}^{\frac NT\le(g(z)+g(z')-\ell\ri)}}{N^{\gamma(2\varkappa-1)} }\frac{P_K(\zeta)P_{K-1}(\zeta')-P_K(\zeta')P_{K-1}(\zeta)}{\eta_{K-1}(\zeta-\zeta')}\big(1+{\cal O}(N^{2\gamma|\delta|-\gamma})\big)\qquad \mbox{when } ~~~|\delta|\neq \frac12. 
\eea
This implies that the statistics of eigenvalues near the hard--edge in the scaling parameter $\zeta$ is governed by the microscopic  (finite--size) matrix model of ``Laguerre'' type.
\subsection{Case $|\varkappa-K|= \frac12$}

When $\varkappa$ is a half-integer, i.e. when $|\delta|=\frac12$, the $\zeta$-independent error term (underbraced part) of (\ref{wtYapprox}) becomes ${\cal O}(1)$.
The following fact is useful:
\be
\frac{M_{K,\delta}}{(\wt C_0 N)^{\gamma}}= \left\{\begin{array}{l}\left[\begin{array}{cc}
{\cal O}(N^{-\gamma}) & \ds -\frac{\eta_K}{2i\pi}\wt C_0^{-\gamma (2K+1)}  \\
{\cal O}(N^{-2\gamma}) & {\cal O}(N^{-\gamma})   \end{array}\right]=-u_K\left[\begin{array}{cc} 0 &  1 \\0  & 0     \end{array}\right]+{\cal O}(N^{-\gamma})
\quad\mbox{when $\delta=\frac12$,}\vspace{0.3cm}\\
 \left[\begin{array}{cc}
 {\cal O}(N^{-\gamma}) &{\cal O}(N^{-2\gamma}) \\
 \ds-\frac{2i\pi}{\eta_{K-1}} \wt C_0^{\gamma (2K-1)}& {\cal O}(N^{-\gamma})  \end{array}\right]=-\ell_{K-1}\left[\begin{array}{cc} 0 &  0 \\1  & 0     \end{array}\right]+{\cal O}(N^{-\gamma})\quad\mbox{when $\delta=-\frac12$,}
 \end{array}\right.
\ee
where we have defined (similar to (4.10) in \cite{BertoLee1})
\be
u_K:=\frac{\eta_K}{2i\pi}\wt C_0^{-\gamma(2K+1)},\qquad \ell_{K-1}:=\frac{2i\pi}{\eta_{K-1}}\wt C_0^{\gamma(2K-1)}.
\ee
By explicitly writing the components as
\be\label{ABcomp}
[\A^{(j)}_K,\B^{(j)}_K]=\left[\begin{array}{cc}
 \A^{(j)}_1 & \B^{(j)}_1  \\
\A^{(j)}_2  & \B^{(j)}_2     \end{array}\right],\ee
the leading behavior of $F_{K,\pm\frac 12}$ (\ref{F1}) is written as
\bea
 F_{K,\frac12}=\frac{u_K}{1+u_K\det[\A^{(0)}_K,\A^{(1)}_K]}
 \left[\begin{array}{cc} \A^{(0)}_1 & 0  \\ 0&     \A^{(0)}_2  \end{array}\right]
  \left[\begin{array}{cc}
  \A^{(0)}_2 & - \A^{(0)}_1  \\ \A^{(0)}_2  &- \A^{(0)}_1      \end{array}\right]+ {\cal O}(N^{-\gamma}) ,
  \\F_{K,-\frac12}=-\frac{\ell_{K-1}}{1+\ell_{K-1}\det[\B^{(1)}_K,\B^{(0)}_K]}
 \left[\begin{array}{cc} \B^{(0)}_1 & 0  \\ 0&    \B^{(0)}_2  \end{array}\right]
  \left[\begin{array}{cc}
 \B^{(0)}_2 & -\B^{(0)}_1  \\\B^{(0)}_2 & -\B^{(0)}_1      \end{array}\right]+ {\cal O}(N^{-\gamma}) .
\eea
where the subscripts stand for the entries of the corresponding matrix.

A straightforward calculation gives the first column of $\wt Y_\infty$ as
\bea\label{1/2}
&&\wt C_0^{\gamma K}N^{\gamma K}\left[\begin{array}{c}
(\wt Y_\infty)_{11} \\ (\wt Y_\infty)_{21} \end{array}\right]\approx
\frac{P_K(\zeta)}{1+u_K\det[\A_K^{(0)},\A_K^{(1)}]}\left[\begin{array}{c} \A^{(0)}_1 \\ \A^{(0)}_2  \end{array}\right]
\\&&\qquad -\frac{\ell_{K-1}\wt C_0^{\gamma} }{N^{\gamma}}P_{K-1}(\zeta)\left(
\left[\begin{array}{c} \B^{(0)}_1 \\ \B^{(0)}_2  \end{array}\right]+u_K\left[\begin{array}{c} \A^{(1)}_1 \\ \A^{(1)}_2  \end{array}\right]+
\frac{u_K\det[\B^{(1)}_K,\A^{(0)}_K]+u_K^2\det[\A^{(2)}_K,\A^{(0)}_K]}{1+u_K\det[\A_K^{(0)},\A_K^{(1)}]}\left[\begin{array}{c} \A^{(0)}_1 \\ \A^{(0)}_2  \end{array}\right]\right) 
\cr&&\qquad+\frac{\zeta P_K(\zeta)}{\wt C_0^\gamma N^\gamma}
\left(\left[\begin{array}{c} \A^{(1)}_1 \\ \A^{(1)}_2  \end{array}\right]+\left[\begin{array}{c} \A^{(0)}_1 \\ \A^{(0)}_2  \end{array}\right]\frac{u_K\det[\A^{(2)}_K,\A^{(0)}_K]}{1+u_K\det[\A_K^{(0)},\A_K^{(1)}]}\right)+{\cal O}(N^{-2\gamma})\qquad \mbox{for} \quad \varkappa=K+\frac12, 
\cr&&\label{-1/2}\wt C_0^{\gamma K}N^{\gamma K}\left[\begin{array}{c}
(\wt Y_\infty)_{11} \\ (\wt Y_\infty)_{21} \end{array}\right]\approx
-\frac{\ell_{K-1}\wt C_0^{\gamma} N^{\gamma} P_{K-1}(\zeta)}{1+\ell_{K-1}\det[\B_K^{(1)},\B_K^{(0)}]}\left[\begin{array}{c} \B^{(0)}_1 \\ \B^{(0)}_2  \end{array}\right]
\\&&\qquad+P_{K}(\zeta)\left(
\left[\begin{array}{c} \A^{(0)}_1 \\ \A^{(0)}_2  \end{array}\right]+\ell_{K-1}\left[\begin{array}{c} \B^{(1)}_1 \\ \B^{(1)}_2  \end{array}\right]-
\frac{\ell_{K-1}\det[\A^{(1)}_K,\B^{(0)}_K]+\ell_{K-1}^2\det[\B^{(2)}_K,\B^{(0)}_K]}{1+\ell_{K-1}\det[\B_K^{(1)},\B_K^{(0)}]}\left[\begin{array}{c} \B^{(0)}_1 \\ \B^{(0)}_2  \end{array}\right]\right) 
\cr&&\qquad-\ell_{K-1}\zeta P_{K-1}(\zeta)
\left(\left[\begin{array}{c} \B^{(1)}_1 \\ \B^{(1)}_2  \end{array}\right]-\left[\begin{array}{c} \B^{(0)}_1 \\ \B^{(0)}_2  \end{array}\right]\frac{\ell_{K-1}\det[\B^{(2)}_K,\B^{(0)}_K]}{1+\ell_{K-1}\det[\B^{(1)}_K,\B^{(0)}_K]}\right)+{\cal O}(N^{-2\gamma})\qquad \mbox{for} \quad \varkappa=K-\frac12. \nonumber
\eea
Here, both approximations consist of three terms which---we must note---are each accurate up to multiplicative errors of $1+{\cal O}(N^{-\gamma})$.
A term containing other dependency on $\zeta$---such as $\zeta^2 P_K(\zeta)$, for instance---comes with ${\cal O}(N^{-2\gamma})$ as noted in the formula.  

The above expressions (\ref{1/2})(\ref{-1/2}) can tell the (asymptotic) location of $K$ (or $K-1$ for the latter formula) zeroes. However, these expressions are not enough to determine the stray zero; for this, we need {\em all the subleading terms}.  Or, the location of stray zeroes have to be computed using the {\em exact} improved outer parametrix.

%\red{ Looking at the above expressions, one can see that the leading location of the zeroes is given by $P_{\varkappa-\frac12}(\zeta)$ when $\varkappa$ is a half-integer.An interesting thing is that, as noted in \cite{BertoLee1}, there appears a stray zero at $\sim N^{\gamma}$ in the $\zeta$-coordinate (i.e. in the finite distance from the port).
%\bea
%\zeta_{\mbox{\scriptsize stray}}&=&\frac{-\wt C_0^{\gamma}N^\gamma \A^{(0)}_1}{\A^{(1)}_1(1+u_K\det[\A^{(0)}_K,\A^{(1)}_K])+u_K\A^{(0)}_1\det[\A^{(2)}_K,\A^{(0)}_K]}+{\cal O}(N^0)\qquad \mbox{for}  \quad \varkappa=K+\frac12,
%\\\zeta_{\mbox{\scriptsize stray}}&=&\frac{\ell_{K-1} \wt C_0^{\gamma}N^\gamma \B^{(0)}_1}{\A^{(0)}_1(1+\ell_{K-1}\det[\B^{(1)}_K,\B^{(0)}_K])-\ell_{K-1}\B^{(0)}_1\det[\A^{(1)}_K,\B^{(0)}_K]}+{\cal O}(N^0)\qquad \mbox{for}  \quad \varkappa=K-\frac12.
%\eea
%Note that the first term is of order $N^\gamma$ in the $\zeta$ coordinate and hence has a limit in the $x$--plane. Of course the actual constant may fall outside the domain of biholomorphicity of $\zeta(x)$ and hence the actual location of these stray zeroes may have to be computed using the improved outer parametrix instead.}

Finally, using (\ref{1/2})(\ref{-1/2}) we write the kernel at $\varkappa=K\pm \frac12$.
\bea 
K_n(\zeta,\zeta')\approx \frac{ \wt C_0^{\gamma}{\rm e}^{\frac NT (g(z)+g(z')-\ell)}}{N^{2\gamma K}}\left(\frac{P_K(\zeta)P_{K-1}(\zeta')-P_K(\zeta')P_{K-1}(\zeta)}{\eta_{K-1}(\zeta-\zeta')}+\frac{\alpha_K}{\eta_K} P_K(\zeta)P_{K}(\zeta')\right)
\\\qquad \alpha_K:=\frac{u_K\det[\A^{(0)}_K,\A^{(1)}_K]}{1+u_K\det[\A^{(0)}_K,\A^{(1)}_K]}\qquad \mbox{for } ~~~\varkappa=K+{\textstyle\frac12},\qquad
\\ 
K_n(\zeta,\zeta')\approx \frac{\wt C_0^{\gamma}{\rm e}^{\frac NT (g(z)+g(z')-\ell)}}{N^{2\gamma (K-1)}}\left(\frac{P_K(\zeta)P_{K-1}(\zeta')-P_K(\zeta')P_{K-1}(\zeta)}{\eta_{K-1}(\zeta-\zeta')}-\frac{\beta_{K-1}}{\eta_{K-1}} P_{K-1}(\zeta)P_{K-1}(\zeta')\right)
\\\qquad \beta_{K-1}:=\frac{\ell_{K-1}\det[\B^{(1)}_K,\B^{(0)}_K]}{1+\ell_{K-1}\det[\B^{(1)}_K,\B^{(0)}_K]}\qquad \mbox{for } ~~~\varkappa=K-{\textstyle\frac12}.\qquad
\eea
These can be considered as ``transitional" kernels when the size of the microscopic matrix model is a half-integer.
Though these kernels do not seem to come from any simple microscopic matrix model, their structures suggest very natural interpretation: a mixture (sum) of two adjacent kernels.
Such behavior of the kernel has been already observed in \cite{BertoLee1}.

We conclude this section with the remark that the above analysis requires non-vanishing denominators $1+u_K\det[\A^{(0)}_K,\A^{(1)}_K]$ and $1+\ell_{K-1}\det[\B^{(1)}_K,\B^{(0)}_K]$. Using the explicit representation (\ref{outerK}) we check (below) that the denominator does not vanish at least for genus 0.
\bea
u_K\det[\A^{(0)}_K,\A^{(1)}_K]=\frac{\eta_{K}4^{1+2K+\alpha}t_0^{2r}}{2\pi\wt C_0^{\gamma(2K+1)}(b-a)^{\alpha}} \left(1-\frac{1}{t_0}\right)^{-2\alpha}\left(\frac{t_0^2}{(b-a)(1-t_0^2)^2}\right)^{1+2K}>0\qquad \mbox{for } ~~~\varkappa=K+{\textstyle\frac12}, 
\\
\ell_{K-1}\det[\B^{(1)}_K,\B^{(0)}_K]=\frac{2\pi\,(b-a)^{\alpha} \wt C_0^{\gamma(2K-1)}}{4^{-1+2K+\alpha}\eta_{K-1}t_0^{2r}} \left(1-\frac{1}{t_0}\right)^{2\alpha}\left(\frac{t_0^2}{(b-a)(1-t_0^2)^2}\right)^{1-2K}>0\qquad \mbox{for } ~~~\varkappa=K-{\textstyle\frac12}. 
\eea

\appendix
\renewcommand{\theequation}{\Alph{section}.\arabic{equation}}

\section{Szeg\"o function in arbitrary genus}\label{App.Szego}

We only sketch the construction since the details would require a good deal of notations to be set up. We will use the same notations and ideas contained in \cite{BertolaMo}.

We denote by $w$ double-cover of the $z$-plane branched at the endpoints of the support of the equilibrium measure
\be
w^2:= \prod_{j=1}^{2g+2} (z-\alpha_j)
\ee
This is a hyperelliptic algebraic curve of genus $g$. We denote by $\infty_{\pm}$ the two points above $z=\infty$ in the usual compactification of the curve, and by $p_{\pm}$ the two points projecting to the location of the hard-edge\footnote{The point $\infty_+$ is characterized by $w>0$ as $z\in \R_+$ near $\infty$. The point $p_+$ is the point on the Riemann surface of $w$ obtained by analytic continuation of $w$ on the complex plane slit along $(\alpha_{2j-1},\a_{2j})$}.
We denote by $\omega_j$ the first-kind differentials normalized along the $a$-cycles: explicitly
\be
\omega_j(z) = \sigma_{j\ell} \frac {z^{\ell-1} \d z}{w}\ ,
\ee
where the summation over repeated indices is understood (and they range from $1$ to $g$) and $\sigma_{j\ell}$ is an invertible matrix such that $\oint_{a_k} \omega_j = \delta_{jk}$.

The notation (rather standard) is lifted from \cite{BertolaMo} and \cite{Faybook}: the Abel map is understood when writing points as arguments of $\Theta$ and it is based at one of the Weierstrass points (for example $\a_1$) $\Delta$ is an arbitrary odd non-singular half-period.
Recall that (pag. 23 of \cite{Faybook}) all such  characteristics $\Delta$ are in one-to-one correspondence with partitions of the Weierstrass points into $g-1$ and $g+3$ points $\{\a_{k_1},\dots, \a_{k_{g-1}} \}\sqcup\{\a_{\ov k_1} ,\dots, \a_{\ov k_{g+3}} \}$

Moreover, straightforward computations show that  (with some {\em overall } ambiguity of signs)
\bea
 \Theta_{_\Delta}(p-\infty_\pm ) \mathop{\sim}_{p\to \infty_{\pm}} \mp \frac 1 z\pa_\ell \Theta_{_\Delta}(0) \sigma_{\ell g}
\eea

The Szeg\"o\ function we want to define now is a  generalization of RHP (\ref{SzegoRHP1})(\ref{SzegoRHP2})(\ref{SzegoRHP3}).
Let
\be
v(z) := \prod_{j=1}^{m} (z-c_j)^{2\rho_j}
\ee
where $\rho_j$ are some real numbers and the points $c_j\in \R$ do not belong to the spectral bands. This function is defined on a simply connected domain obtained by removing (for example) vertical half-lines in the upper half--plane.
The Szeg\"o\ function is characterized by the following scalar Riemann--Hilbert problem
\bea\label{SzegoRHP1b}
&&D_+(z)D_-(z)  = v(z), \ \hbox{ for $z$ on the cut(s)}, \\
&&\lim_{z\rightarrow\infty}D(z) = D_\infty \quad\mbox{where } D_\infty\in{\mathbb R}_+,\label{SzegoRHP2b}
\\&& D(x) \mbox{ is analytic and non-zero otherwise}. \label{SzegoRHP3b}
\eea
We denote by $\infty_{\pm},\pm \gamma_j$ the points on the hyperelliptic curve above $z=\infty,z=c_j$ on the two sheets of the covering  defined in the customary way; we will refer to the sheet of the points $\infty_+, \gamma_j$'s as the {\em physical sheet}.
We leave it to the reader to verify that the solution can be written as follows
\be D(z)= \prod_{j=1}^m\left(\frac{\Theta_\Delta(p-\infty_-)\Theta_\Delta(p+\gamma_j)}{\Theta_\Delta(p-\infty_+)\Theta_\Delta(p-\gamma_j)}\right)^{\rho_j} v(z)^{1/2}.
\ee
The first and third  conditions (\ref{SzegoRHP1b})(\ref{SzegoRHP3b}) are easily checked.
% while the third one (\ref{SzegoRHP3}) follows from the fact that the product of Theta functions has singularities only above $c_j$ and $\infty$, which cancel precisely the singularities of $\omega^{\frac 12}$ on the  physical sheet.
The second condition (\ref{SzegoRHP2b}) can be verified by taking the limit of the proposed expression
\be
\lim_{z=\infty} D(z) =\le(\frac {\Theta_\Delta(\infty_+- \infty_-)}{ \pa_\ell \Theta_{_\Delta}(0) \sigma_{\ell g}}\ri)^{\sum \rho_j}
 \prod_{j=1}^{m} \frac {\Theta_{\Delta}(\infty_+ + \gamma_j) }{\Theta_{\Delta}(\infty_+ - \gamma_j)}
\ee
Note that -since all branchpoints are real- the Theta-function  $\Theta_\Delta(p-\infty_+)$ is real for any  $p\in \R$ and outside of the spectral bands as follows from (\cite{Faybook}, Chapter VI, Prop. 6.1). So is the matrix $\sigma_{kl}$ and the derivatives of $\Theta_\Delta$ at the origin. Hence the expression above is real and -in case it is negative- we can simply change the overall sign to  the expression for $D$.

The expression above needs to be supplemented by specifying where the cuts are made; for the denominator the cuts run on the physical sheet between $\gamma_j$ and $\infty_+$, while for the numerator they run similarly but on the second sheet.  It is simple to verify (due to the fact that $\Theta_\Delta(p-q)$ has a simple zero for $p=q$) that in fact the analytic continuation of $D(x)$ around each $c_j$ yields the same germ of holomorphic function and hence $D(x)$ has --in fact-- no cuts on the physical sheet.

\section{Partial Schlesinger transform and any order asymptotics}\label{App.PST}

Here we present a procedure to improve the error bound of the asymptotic solution to an arbitrary accuracy.  All-order-asymptotics has appeared in \cite{ErcolaniMclaughlin02}.  There, the improved asymptotics are modified by a right multiplication to the strong asymptotics.  This right multiplication can render the jump matrix become as close to the identity as one wishes (since the jump also acts to the right) by a certain well-defined recursive procedure, hence enabling one to get a solution with any improved error bound.

Our method is a little different in that we use a {\em  left} multiplication on both outer parametrix and local parametrix.  In the main text, we have already used such method without which the asymptotics would have failed at any half-integer $\varkappa$.
Here we explain how this method can be applied to achieve an arbitrarily high order of accuracy in the error commited when matching the outer parametrix to the local parametrix at the hard-edge. It should be said that there is no practical purpose in improving the error that comes from the hard--edge (or outpost) parametrix to better than $\mathcal O(N^{-1})$ since then errors coming in the matching between the local parametrices at the other (simple) turning points are  already of that order. In principle the same idea could be applied to the endpoints, but the details become different and in the interest of conciseness we refrain from such analysis.

Inside the disk ${\mathbb D}$, the strong asymptotics of $\wt Y$ is the product of the global parametrix and the local parametrix, i.e. $\wt Y_\infty=\Psi_\varkappa R_\varkappa$.  Outside the disk ${\mathbb D}$ the strong asymptotics is simply $\wt Y_\infty=\Psi_\varkappa$ and, therefore, we have a jump $R_\varkappa$ on the boundary $\partial {\mathbb D}$.

To have a good asymptotic solution, this jump must be as close to the identity as possible.   To this effect we will ``improve" $R_\varkappa$\footnote{We will use the same notations $R_\varkappa$ for the ``improved" local parametrix and $\Psi_\varkappa$ for the ``improved" global parametrix.} by a suitable left multiplication.  This left multiplication will introduce  a new singularity.  Since the strong asymptotics $\Psi_\varkappa R_\varkappa$ must be holomorphic at the outpost, $\Psi_\varkappa$ also needs to be modified by a {\em left} multiplication (that resembles a Schlesinger transformation) so as to cancel the singularity.  Below we describe the whole procedure.

Using (\ref{RK}) $R_K:=\wt z^{-K\sigma_3} H_K=\zeta^{-K\sigma_3}\big({\bf 1}+{\cal O}(1/\zeta)\big)\, \zeta^{K\sigma_3}$ we can obtain the following expansion.
\begin{equation}\label{zRz}
\wt z^{K\sigma_3}R_K \wt z^{-K\sigma_3}={\bf 1}+\frac{Y_1}{\wt z}+\frac{Y_2}{\wt z^2}+\cdots,
\end{equation}
 where $Y_j$'s are matrices of order ${\cal O}(N^{-\gamma j})$ from the consideration of $\wt z\sim N^{-\gamma} \zeta$.  This grading property is crucial for the error analysis.

 Since $\det H_K\equiv 1$ both sides of the above equation (\ref{zRz}) are also uni-modular.
In such case, we claim that the following factorization can be done
\bea\label{zRz}
\wt z^{K\sigma_3}R_K \wt z^{-K\sigma_3}=\left({\bf 1}+\frac{M_1}{\wt z}\right)\left({\bf 1}+\frac{\wt M_1}{\wt z}\right)\left({\bf 1}+\frac{M_2}{\wt z^2}\right)\left({\bf 1}+\frac{\wt M_2}{\wt z^2}\right)\cdots\left({\bf 1}+\frac{M_p}{\wt z^p}\right)\left({\bf 1}+\frac{\wt M_p}{\wt z^p}\right)\times
\\\times\left({\bf 1}+{\cal O}\left(\frac{1}{\wt z^{p+1}}\right)\right)\qquad \mbox{for arbitrarily large positive integer $p$}, 
\eea
where $M_j$s ($\wt M_j$s) are all nilpotent matrices so that each factor is uni-modular.

The proof proceeds by induction.  First, any uni-modular matrix that has an expansion  $  {\bf 1}+\frac{A}{\wt z^{p+1}}+{\cal O}\left(\frac{1}{\wt z^{p+2}}\right)$ must have a traceless matrix $A$.  Secondly, any 2 by 2 traceless matrix can be represented as  the  sum of two nilpotent matrices, say $A= M_{p+1}+\wt M_{p+1}$.  Then the series can be expressed as
\be{\bf 1}+\frac{A}{\wt z^{p+1}}+{\cal O}\left(\frac{1}{\wt z^{p+2}}\right)=\left({\bf 1}+\frac{M_{p+1}}{\wt z^{ p+1}}\right)\left({\bf 1}+\frac{\wt M_{p+1}}{\wt z^{ p+1}} \right)\left({\bf 1}+{\cal O}\left(\frac{1}{\wt z^{p+2}}\right)\right).\ee
This gives the recursive step of increasing $p$ by one in (\ref{zRz}). {\bf Q.E.D.}

With such expansion in mind, we define the improved local parametrix by
\be R_\varkappa:=\wt z^{-K\sigma_3}\left({\bf 1}-\frac{M_1}{\wt z}\right)\left({\bf 1}-\frac{\wt M_1}{\wt z}\right)\cdots\left({\bf 1}-\frac{M_p}{\wt z^p}\right)\left({\bf 1}-\frac{\wt M_p}{\wt z^p}\right)\,\wt z^{K\sigma_3} R_K.\ee
Here we have used $\left({\bf 1}+\frac{M}{\wt z}\right)^{-1}={\bf 1}-\frac{M}{\wt z}$ for any nilpotent $M$.  Using the grading property that we mentioned earlier, we obtain $R_\varkappa={\bf 1}+{\cal O}(N^{-\gamma(p+1)})$, which will be the error bound of our improved asymptotics.

Now we improve the global parametrix such that $\Psi_\varkappa R_\varkappa$ is holomorphic inside ${\mathbb D}$. We recall that $\Psi_K$ has a singularity at the outpost such that $\Psi_K \wt z^{-K\sigma_3}$ is holomorphic at the outpost.
We propose that
\bea \label{appPsi} \Psi_\varkappa =F(z)\, \Psi_K=\wt F_p(z)F_p(z) \cdots \wt F_2(z)F_2(z)\,\wt F_1(z)F_1(z)\,\Psi_K,
\\ F_j(z):={\bf 1}+\frac{F_{j1}}{z}+\frac{F_{j2}}{z^2}+\cdots++\frac{F_{jj}}{z^j}\qquad (\mbox{and similarly for $\wt F_j(z)$}).
\eea

The (recursive) construction follows.
To illustrate the method let us write out the asymptotics.
\be\Psi_\varkappa R_\varkappa=\cdots\times\overbrace{ \left({\bf 1}+\frac{\wt F_{11}}{z}\right)\underbrace{\left({\bf 1}+\frac{F_{11}}{z}\right) \Psi_K \wt z^{-K\sigma_3} \left({\bf 1}-\frac{M_1}{\wt z}\right)}_{\mbox{\small part }1}\left({\bf 1}+\frac{\wt M_{1}}{\wt z}\right)}^{\mbox{\small part }2} \times\cdots. \label{part1part2}\ee
We understand $\Psi_K\wt z^{-K\sigma_3}$ is holomorphic at the outpost. First we will find $F_{11}$ that makes ``part 1"  holomorphic at the outpost.  Having a holomorphic ``part 1", we will then find $\wt F_{11}$ that makes ``part 2'' holomorphic.

Obviously, these steps are recursive and we only need to clarify one generic step: finding $j$ matrices, $F_{j1},\cdots,F_{jj}$, that satisfy the following analyticity condition.  
\be \left({\bf 1}+\frac{F_{j1}}{z}+\cdots+\frac{F_{jj}}{z^j}\right)\, \A(z)\,\left({\bf 1}-\frac{M_j}{\wt z^j}\right)={\cal O}(z^0),\label{Fjjcond}
\ee
where $\A(z)$ is a matrix whose entries are analytic at the hard-edge (or outpost) which is determined from the previous step in the induction process (e.g. part 1 and part 2 in (\ref{part1part2})). 
Denoting the expansion of $\A$ as $\A(z)=A_0+A_1 z+A_2 z^2+\cdots,$ and defining a new expansion:
\be \A^{[j]}:=\A(z)\left(\frac{z}{\wt z}\right)^j=A^{[j]}_0+A^{[j]}_1 z+A^{[j]}_2 z^2+\cdots,\ee
we can write the following solution:
\bea 
&&(F_{j1},\cdots,F_{jj})=
\\&&\quad=(A^{[j]}_0,\cdots,A^{[j]}_{j-1})\cdot M_j\cdot
\left(\left[\begin{array}{cccc}0&\cdots&0&A_0\\0&\cdots&A_0&A_1\\\vdots&&&\vdots\\A_0&A_1&\cdots&A_{j-1}\end{array}\right]-\left[\begin{array}{cccc}A^{[j]}_1&A^{[j]}_2&\cdots&A^{[j]}_j\\A^{[j]}_2&A^{[j]}_3&\cdots&A^{[j]}_{j+1}\\\vdots&&&\vdots\\A^{[j]}_j&A^{[j]}_{j+1}&\cdots&A^{[j]}_{2j-1}\end{array}\right]\cdot M_j\right)^{-1}.\eea
The global parametrix $\Psi_\varkappa$ (\ref{appPsi}) is then fully determined by the recursive procedure.

\section{Construction of the outer parametrix for arbitrary number of cuts}
\label{multicut}
We refer to the corresponding appendix in \cite{BertoLee1} since the formul\ae\  are identical, the only difference being in the picture that accompanies the appendix.

\bibliographystyle{plain}
\bibliography{PortColonization_2}
\end{document}